\documentclass{egpubl}
\usepackage{egsr2022_DLOnlyTrack}
\usepackage{xcolor}
 
 \WsPaper           

\usepackage[T1]{fontenc}
\usepackage{dfadobe}  
\usepackage{amssymb}

\biberVersion
\BibtexOrBiblatex
\usepackage[backend=biber,bibstyle=EG,citestyle=alphabetic,backref=true]{biblatex} 
\electronicVersion
\PrintedOrElectronic

\ifpdf \usepackage[pdftex]{graphicx} \pdfcompresslevel=9
\else \usepackage[dvips]{graphicx} \fi

\usepackage{egweblnk} 
\usepackage[percent]{overpic}

\newcommand{\eqref}[1]{(\ref{#1})}

\newcommand\new[1]{{\color{black}#1}}

\newcommand{\Real}{\mathbb{R}}

\newcommand{\px}{\mathbf{x}}
\newcommand{\Ray}{\mathrm{r}}

\newcommand{\Lo}{L_o}
\newcommand{\Li}{L_i}
\newcommand{\Le}{L_e}

\newcommand{\T}{\mathrm{T}}

\newcommand{\SH}[1]{\bar{#1}}

\newcommand{\Dataset}[1]{\textsc{#1}}

\title[A Learned Radiance-Field Representation for Complex Luminaires]%
      {A Learned Radiance-Field Representation\\for Complex Luminaires}

\author[Condor \&  Jarabo]
{\parbox{\textwidth}{\centering Jorge Condor$^{1,2}$\orcid{0000-0002-9958-0118}
       and Adrian Jarabo$^{1}$\orcid{0000-0001-9000-0466} } \\
{\parbox{\textwidth}{\centering $^1$Universidad de Zaragoza - I3A, Spain\\ $^2$Università della Svizzera Italiana, Switzerland } } }

\begin{document}

\teaser{
  \includegraphics[width=0.33\linewidth]{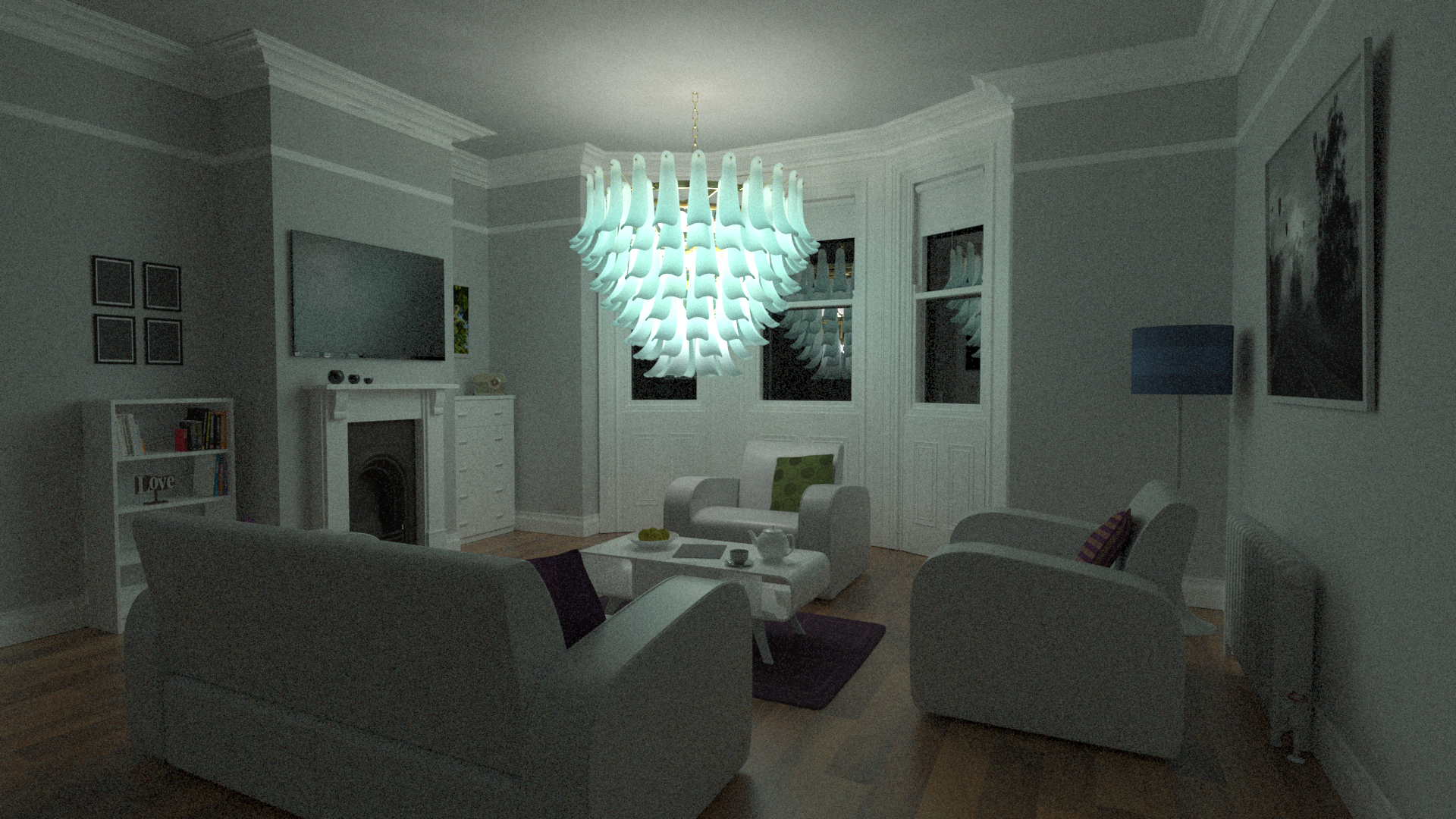}
 \includegraphics[width=0.33\linewidth]{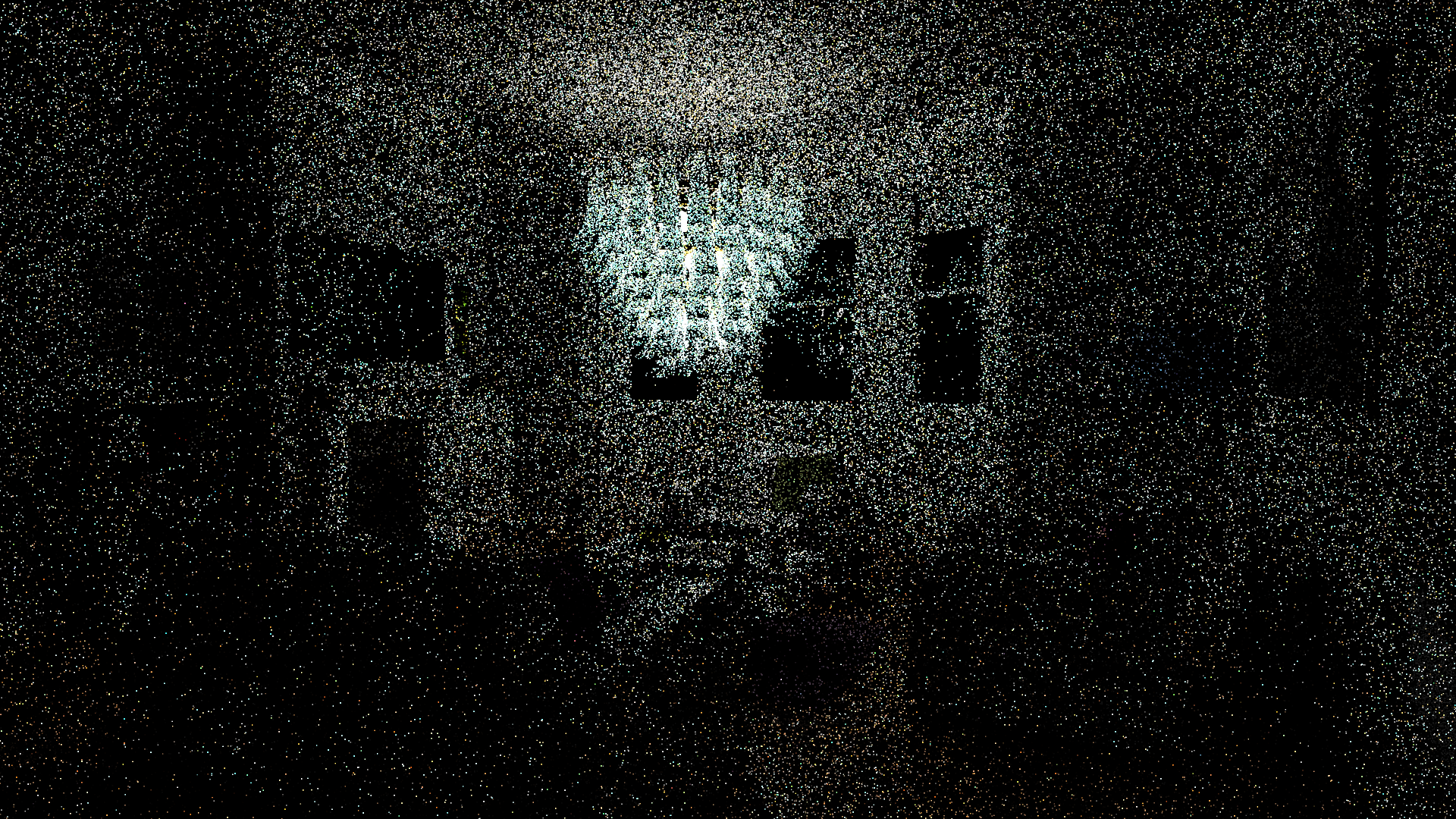}
 \includegraphics[width=0.33\linewidth]{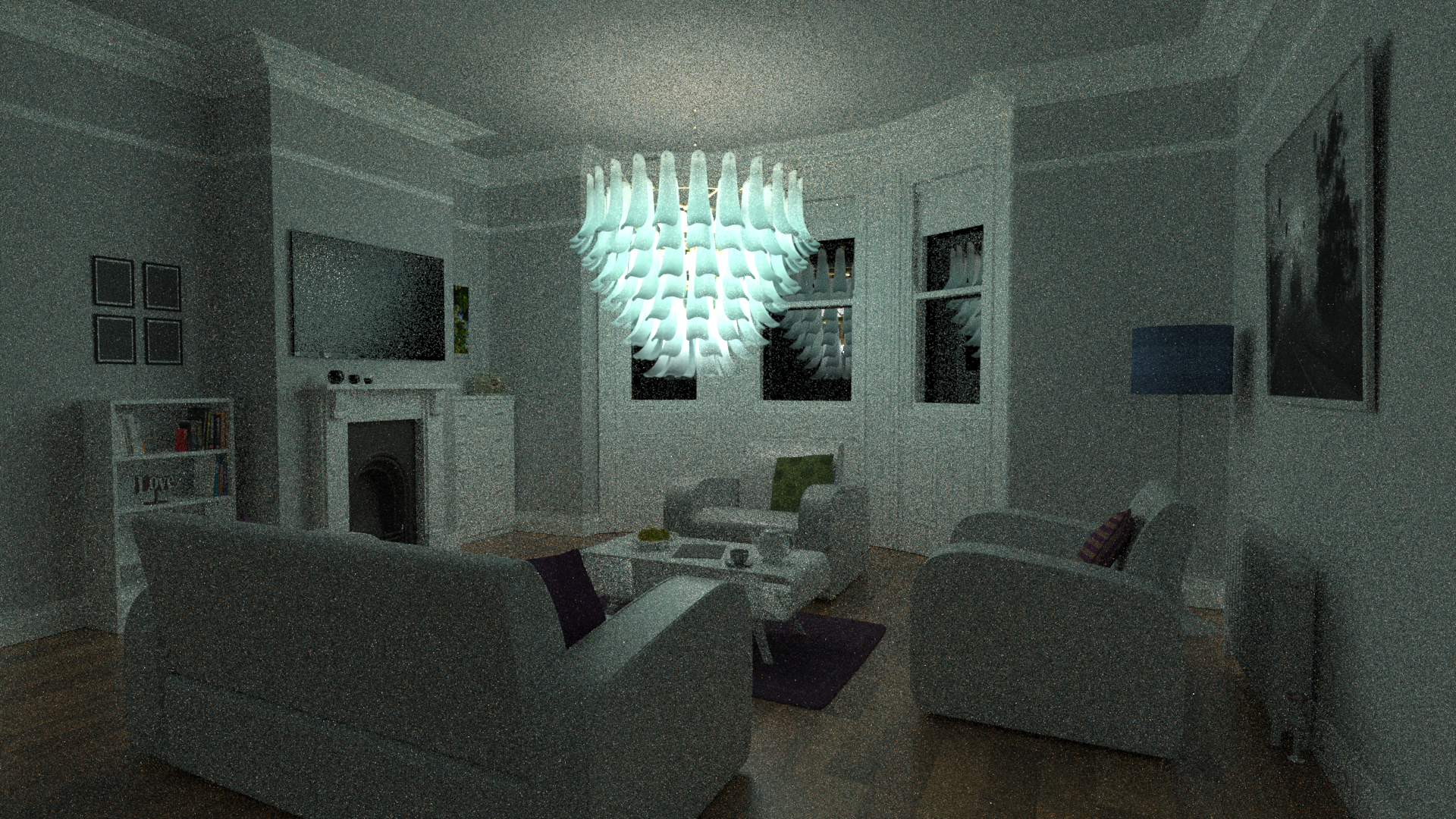}
 \centering
  \caption{\textbf{A living room scene containing a learned radiance field-based complex luminaire.} From left to right: Reference path-traced scene with explicitly-modeled complex luminaire (32768 samples per pixel, 7.2h); the same scene rendered with 64 samples per pixel in 1.3 minutes; and our method with 32 samples per pixel, 52.1 seconds. We leverage learned volumetric radiance fields to obtain high quality representations of complex luminaires.}
\label{fig:teaser}
}

\maketitle
\begin{abstract}
   We propose an efficient method for rendering complex luminaires using a high quality octree-based representation of the luminaire emission. 
   Complex luminaires are a particularly challenging problem in rendering, due to their caustic light paths inside the luminaire. 
   We reduce the geometric complexity of luminaires by using a simple proxy geometry, and encode the visually-complex emitted light field by using a neural radiance field. We tackle the multiple challenges of using NeRFs for representing luminaires, including their high dynamic range, high-frequency content and null-emission areas, by proposing a specialized loss function. 
   For rendering, we distill our luminaires' NeRF into a plenoctree, which we can be easily integrated into traditional rendering systems. Our approach allows for speed-ups of up to 2 orders of magnitude in scenes containing complex luminaires introducing minimal error.

\ccsdesc[300]{Computer graphics~Neural Rendering}
\ccsdesc[300]{Machine Learning~Neural Radiance Fields}

\printccsdesc   
\end{abstract}  

\section{Introduction}
\label{sec:intro}


Complex luminaires are ubiquitous in real-world scenes, from isolated light bulbs with tiny coils embedded in dielectric bulbs, to complex chandeliers made by thousands of small glass pieces that scatter the light of the (potentially many) emitters enclosed. While a common approach is to reduce the complexity of light sources to simple distant or area lights that can be compactly evaluated and sampled~\cite{bitterli2015portal,guillen2017area,peters2021brdf}, in many scenarios including architectural visualization such simplifications result into a significant loss in realism. However, as the light source increases in complexity these approaches become unfeasible: the complex light paths connecting the potentially multiple individual light sources in the luminaire, and the surfaces being illuminated by simply random chance is unlikely in the best of cases. Thus, computing the illumination from a complex luminaire would require many thousands of samples to converge, even using complex Markov-Chain Monte Carlo methods~\cite{jakob2012manifold}. 

Precomputing the radiance field of the luminaires is an effective approach for avoiding the sampling of difficult paths in run time~\cite{velazquez2015complex,zhu2021neural}. It allows to encode all light paths exiting the luminaire as a five-dimensional proxy function, which is fast to access during rendering: to determine the illumination at a point, it is only needed to integrate over the proxy function. 
Unfortunately, accurately precomputing a luminaire requires a very dense storage of the five-dimensional function modeling the emission of luminaires, which might become unfeasible due to memory constraints, or to combine the precomputed proxy with the real complex geometry of the scene.

In this work we tackle this problem by leveraging the potential of recent advances on volumetric representations of appearance, and in particular on the compact and efficient representation of radiance fields (RFs)~\cite{mildenhall2020,tewari2021advances}.
%
Radiance Fields, including \emph{neural radiance fields} (NeRF), have emerged as a suitable representation of very complex three-dimensional scenes by encoding the scene as a lightweight volumetric function. The key characteristic of this volumetric function, in contrast to previous voxelization-based representations of appearance~\cite{neyret98, loubet2017}, is that this function is learned from a (relatively) sparse set of views. 
RFs allow to compactly encode highly-complex spatio-directional representations of 3D scenes, which can be efficiently rendered by standard ray-marching through the volumetric proxy. 

Unfortunately, off-the-shelf RFs are not adequate for our particular task, and modeling a luminaire using a RF, and integrating it into a Monte Carlo rendering engine, introduces several challenges, including: 1) Luminaires have high dynamic range (HDR), as opposed to the usual low dynamic range (LDR) targeted by NeRFs, and 2) exhibit high angular frequency; and 3) they require handling both opaque and transparent elements for integration in synthetic scenes. 
We solve these challenges using a training loss designed for handling HDR content, and a combination of a linear volumetric representation of the RF inspired in correlated media~\cite{vicini2021non,jarabo2018radiative}. 

We integrate our RF-based luminaire in a physically-based rendering using an efficient octree-based representation of the radiance field, demonstrating speed-ups of up to two orders of magnitude, compared with traditional path tracing and volumetric path tracing methods with little error.

\section{Related Work}
\label{ch:rw}

\paragraph*{Volumetric representations of appearance}
Volumetric approaches have been successfully used to approximate the appearance of complex geometries in a wide array of different applications, from trees ~\cite{loubet2017,neyret98}, cloth and hair ~\cite{aliaga17, khungurn15, shroder11, zhao11}, or particulate media (sugar, salt, etc)~\cite{meng15, moon07, mueller16efficient}. These works rely on computing (potentially heterogeneous) bulk scattering parameters by using anisotropic radiative transfer~\cite{jakob2010radiative,heitz2015sggx}. 
Zhao et al.~\cite{zhao2016downsampling} proposed a method for downsampling this type of volumetric appearances by optimizing their directional scattering. 
Recently, Vicini et al.~\cite{vicini2021non} improved the accuracy of volumetric representations of surface-based scenes by using a non-exponential correlated volumetric representation~\cite{jarabo2018radiative,bitterli2018radiative}, allowing to represent a larger range of media (correlated or otherwise) with a higher level of detail. All these works assume scattering media with no emission. Our work builds upon these ideas, using a volumetric representation for approximating the appearance of complex luminaries. 

\paragraph*{Light and Radiance Fields}
Light fields~\cite{wu2017light} are vector functions that encode the radiance carried by a particular ray defined in a four-dimensional space, and are a compact form of rendering complex scenes based on images. Recently, with the widespread of deep learning methods, a number of works have been proposed to encode these light fields in compact neural-based representations~\cite{neuralVolumes, srn, mildenhall2019llff}. 

Radiance fields (RF)~\cite{mildenhall2020} encode the ray-to-radiance mapping by using a volumetric approach, in which each point of the volume space encodes its directional radiance. In rendering time, a volume rendering-based approach is used, by ray marching the volumetric representation of the scene and gathering the radiance at each point in the ray. This directional radiance is learned by using a set of images of the scene. 
In order to efficiently encode the RF, Mildenhall et al.~\cite{mildenhall2020} proposed to use a neural network (a neural radiance field, or NeRF). Since its inception, NeRFs have received an outstanding attention; we refer to recent surveys on neural rendering for an overview of methods building upon and generalizing NeRF~\cite{tewari2021advances}. Unfortunately, NeRFs require evaluating a neural network for each ray marching step, which might be expensive. Yu et al.~\cite{yu2021plenoctrees} proposed to distill the NeRF into a primal-space octree encoding the directional radiance in spherical harmonics. This allowed them to prune non-contributing empty regions, and perform very fast rendering of surface-based radiance fields. We extend this approach to encode the radiance field of complex luminaires.

\paragraph*{Complex Luminaries} 

The most common approach for efficiently rendering complex luminaries is to approximate the luminaire by a simpler proxy geometry, and encode its complexity by precomputing the near-field outgoing emission from the proxy. 
Early works~\cite{Ashdown1995NearFieldPM,ngai87} tabulated the complex five-dimensional emission function. Heidrich et al.~\cite{Heidrich1998CannedL} proposed to encode the emission using light fields, following an approach similar to image-based rendering techniques
Rayset-based approaches~\cite{rykowski98,mas08,muschaweck2011} follow a similar idea, but removing the structure of the outgoing emission, by capturing the spatio-directional distribution of radiance at the proxy. Lu et al.~\cite{lu2014position} proposed a method for importance sampling light field-based representations of luminaires. All these works focus on modeling the emission of light sources in the scenes, but cannot represent their appearance unless a very high resolution is used to represent the emission.



More recently, Velázquez-Armendáriz et al.~\cite{velazquez2015complex} proposed a hybrid representation of the luminaire, representing its illumination as a set of directional point light sources which can be used for lighting the scene, while its appearance is rendered by using the full geometry of the luminaire and a precomputed volume emission. In contrast, our approach seamlessly models both illumination and appearance. 

Finally, the closest method to ours was proposed by Zhu et al.~\cite{zhu2021neural}, that leverage neural rendering techniques for encoding the appearance and illumination from complex luminaires. They use two multilayer perceptrons (MLPs) to learn the full light field of the luminaire approximated as a simple proxy, as well as its transparency for integration in the synthetic scene. While the results are of high-quality, they require several thousands of input images for training, the results are slightly oversmoothed, and the evaluation is expensive, requiring a dedicated GPU for evaluating the MLPs. Our work is similar in spirit, but instead leverages recent learned volumetric radiance fields for encoding the luminaire, resulting in fast and high-quality rendering of luminaires. It implicitly encodes transparency for matting, requires less training data, and does not need to evaluate a neural network in rendering time.

\section{Problem Statement}
Our goal is to evaluate a complex luminaire as efficiently as possible within any given scene, so that it can be used for both modeling its appearance (e.g. as seen from the camera) and its illumination, making it easy to integrate in any Monte Carlo-based rendering engine. 

\begin{figure*}[!t]
    \includegraphics[width=\textwidth]{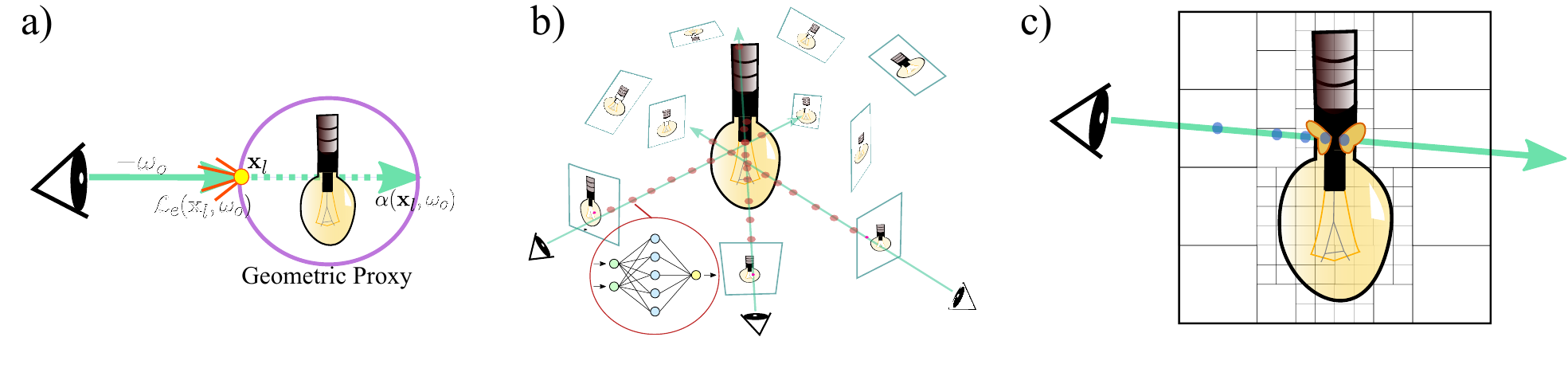}
    \vspace{-1cm}
    \caption[\textbf{Overview:} of the method]{\textbf{Overview: } \new{a) We approximate the appearance of a complex luminaire by using a proxy geometry and encoding the emitted field $\Le(\px,\omega)$ as a five-dimensional function in the proxy surface, plus a transparency function $\alpha(\px,\omega)$ for integrating the luminaire. Previous approaches~\cite{zhu2021neural} directly learn the emitted field and transparency function, requiring thousands of training images to converge. b) We exploit recent advances in neural radiance fields (NeRFs) to first encode the emitted field using a multi-layer perceptron (MLP) using a volumetric rendering approach \cite{mildenhall2020}, using a custom loss function for its training specifically tailored to accurately model the emitted field of complex luminaires. (c) To avoid neural queries on render time, we project the angular domain of the NeRF on spherical harmonics, and the spatial domain in an octree, following Yu et al.\cite{yu2021plenoctrees}. This allows us to significantly limit the spatial queries, leading to a very efficient volumetric representation which is easy to integrate in any render engine. }  
    }
    \label{fig:overview}
\end{figure*}
We start by following previous works on luminaire modeling~\cite{Heidrich1998CannedL,zhu2021neural}: These works partially (or totally) remove the geometric complexity and precompute the light transport within the luminaire. 

We first make the assumption that it will always be rendered from the outside; this is a reasonable assumption for most applications, even for close-up shots, as long as we don't place the camera inside the actual luminaire (i.e. inside the dielectric bulbs). This allows us to simplify the geometry of the luminaire as a simple proxy geometry bounding the luminaire.
Then, the appearance of the luminaire can be encoded in a per-ray basis as a 5D function $\Lo$ modeling the outgoing radiance exiting the proxy at position $\px_l$ in direction $\omega_o$ as 
\begin{equation}
    \Lo: \Real^3 \times \Real^2 \rightarrow \Real; \qquad (\px_l, \omega_o) \mapsto \Lo(\px_l, \omega_o).
    \label{eq:f_mapping}
\end{equation}
In order to model $\Lo(\px_l, \omega_o)$ we need to account for two terms: The emitted field of the luminaire $\Le(\px_l,\omega_o)$, that encodes all light paths starting at the luminaire and outgoing in $(\px_l, \omega_o)$, and the light from the scene that might pass through the luminaire (due to empty space within the proxy), which is crucial for integrating the luminaire in the scene. Therefore, we model $\Lo(\px_l,\omega_o)$ as
\begin{equation}
    \Lo(\px_l,\omega_o) = \Le(\px_l,\omega_o) + \alpha(\px_l,\omega_o) \, \Li(\px_l-\omega_o\cdot t_1,\omega_o),
    \label{eq:ours}
\end{equation}
with $\alpha(\px_l,\omega_o)$ the transparency along a ray with origin at $\px_l$ and direction $-\omega_o$. $\Li(\px,\omega)$ is the incoming radiance at the proxy in point $\px$ from direction $\omega$, and $t_1$ is the intersection distance of the ray with origin at $\px_l$ and direction $-\omega_o$ with the proxy geometry. This approach is illustrated in Figure~\ref{fig:overview} (1). 

Thus, the key problem is how to model the emission and transparency functions $\Le$ and $\alpha$ respectively, such that a) they can be efficiently evaluated in rendering time, b) they are compact from a storage perspective, c) preserve their spatio-directional high frequency, and d) can deal with the potentially large dynamic range of the luminaires. Zhu et al.~\cite{zhu2021neural} proposed to use two specialized image-based neural rendering systems for each of these terms, at the cost of requiring a large training set. 

However, we observe that modern volumetric-based \emph{radiance fields} (RF) excel at learning such 5D mapping (ray to radiance), even in very complex scenes with a relatively small training set. Moreover, since RF are based on volume rendering, they implicitly encode the transparency term $\alpha(\px_l,\omega_o)$. Therefore it is natural to pose our problem by using a neural representation of the luminaire radiance field.

\paragraph*{Challenges} Unfortunately, directly applying a state-of-the-art RF to model synthetic complex luminaires for rendering does not provide good results, and there are several challenges that need to be addressed:

%

\begin{itemize}
\item \textbf{High Dynamic Range}. Learned radiance fields have mainly focused on capturing LDR scenes. However, in a physically-based rendering engine, we need to account for high-dynamic-range radiance. Extending these methods to work on HDR is challenging, since the areas with the highest radiance will dominate the gradient and thus the training. 

\item \textbf{Null Emission}. Since we are modeling isolated objects that will be integrated into synthetic scenes, we need to account for zero-radiance light rays. In contrast, in natural scenes these absolute-black areas are uncommon. In a general machine learning context, dark pixels generate very small gradients during training, which makes them a difficult training target as the error is dominated by the more radiant parts. 

\item \textbf{Matting}. In order to blend our luminaire with the rest of the scene, we will need to let through the proxy those rays that actually hit the convex geometry but not the underlying light source. This requires us to incorporate not only the emission from a ray, but also the amount of light that passes through the proxy of the luminaire (i.e. the \emph{transparency}). A highly accurate transparency (or opacity) model is necessary for a good integration of the luminaire within the scene, which sometimes NeRF struggles with.
\end{itemize}

We address these issues by introducing a loss function targeting the particularities of luminaires' radiance fields. In the following, we detail our model. 
\section{Our Model}

As discussed earlier, our approach leverages recent works on learned volumetric radiance fields for modeling  the appearance and illumination of luminaires. These are learned from a sparse set of HDR training views around the luminaire. These radiance fields are then encoded into an efficient octree-based structure, that allows us to efficiently query our luminaires during render time. In the following, we detail the different elements of our model.

\subsection{Physical Model}
Volumetric radiance fields~\cite{mildenhall2019llff} represent the scene's spatio-directional radiance (light) field as the volume density and directional emitted radiance at any point in space, by using traditional volume rendering techniques. For each outgoing point and direction $\px$ and $\omega$, we define a ray $\Ray(t)=\px - \omega \, t$, and compute the emitted field of the luminaire $\Le(\Ray):=\Le(\px,\omega)$ as
\begin{equation}
    \label{eq:nerfequation}
     \Le(\Ray) = \int_{0}^{t_1}\T(\Ray,t) \, \sigma\left(\Ray(t)\right) \, \Phi \left(\Ray(t), \omega\right)\, dt, 
\end{equation}
with $t_1$ farthest intersection distance with the proxy geometry, $\sigma\left(\px\right)$ and $\Phi\left(\px, \omega\right)$ are the density and emission at point $\px$ towards direction $\omega$, and $\T(\Ray,t)$ is the transmittance. 

While transmittance is in general modeled using the classic exponential Beer-Lambert law, Vicini et al.~\cite{vicini2021non} showed that non-exponential light transport~\cite{jarabo2018radiative,bitterli2018radiative} is a better-suited model for radiance fields, specially when encoding surfaces due to their implicit correlation. Therefore, we opt for Vicini's linear transmittance, defined as
\begin{equation}
    \label{eq:transmittance}
    \T(\Ray,t) = \max\left(0, 1 - \int_{0}^{t} \sigma\left(\Ray(s)\right) \, ds\right).
\end{equation}
Finally, we model the transparency $\alpha(\Ray):=\alpha(\px,\omega)$ as
\begin{equation}
    \alpha(\Ray) = 1 - T(\Ray,t_1). 
    \label{eq:alpha}
\end{equation}

We numerically approximate Equations~\eqref{eq:nerfequation} and \eqref{eq:transmittance} by deterministically ray-marching the ray inside the proxy geometry, following the standard procedure in NeRF and other RF-related papers. This operation is fully differentiable, which is well-suited for learning the scene from a set of views.

\subsection{Encoding the Radiance Field} As shown in Equation~\eqref{eq:nerfequation}, we model the RF as a heterogeneous medium defined by its density $\sigma(\px)$ and directional emission $\Phi(\px,\omega)$. The vast body of works on learned-RF representations use a neural network for jointly modeling $\sigma(\px)$ and $\Phi(\px,\omega)$ (NeRFs); unfortunately, querying a NN might be expensive, and introduces additional complexity in terms of integration with a rendering engine. Instead, we follow the approach by Yu and colleagues~\cite{yu2021plenoctrees}, and project the trained NeRF into a voxelized octree representation of the radiance field (a \emph{plenoctree}), which is compact, CPU friendly, and easy to integrate in any rendering engine. 

We first train a NeRF that captures the full dynamic range of the light source. The transition from the common LDR to HDR introduces several practical problems in terms of encoding and training. We solve that by using a training loss that introduces a regularizer that copes with the potentially large radiance magnitudes in HDR. In addition, given the importance of a good transparency term $\alpha(\Ray)$, we directly supervise the opacity during training. 

Then, after training, we project the density $\sigma(\px)$ and emission $\Phi(\px)$ in a plenoctree. To ease such projection, instead of learning the whole 5-D spatio-directional function, the neural network learns the spherical harmonics (SH) expansion of $\Phi(\px,\omega)$ defined by the SH coefficients $\SH{\Phi}_l^m(\px)$, with $l$ and $m$ the SH level and band of the coefficient respectively. 

In the following, we describe our neural network architecture, loss function and training details of our NeRF-based luminaires. Then, we discuss the projection to a plenoctree, and how we integrate it into a path tracer.

\begin{figure}[!htbp]
   \begin{minipage}[b]{\columnwidth}
    \includegraphics[width=\columnwidth]{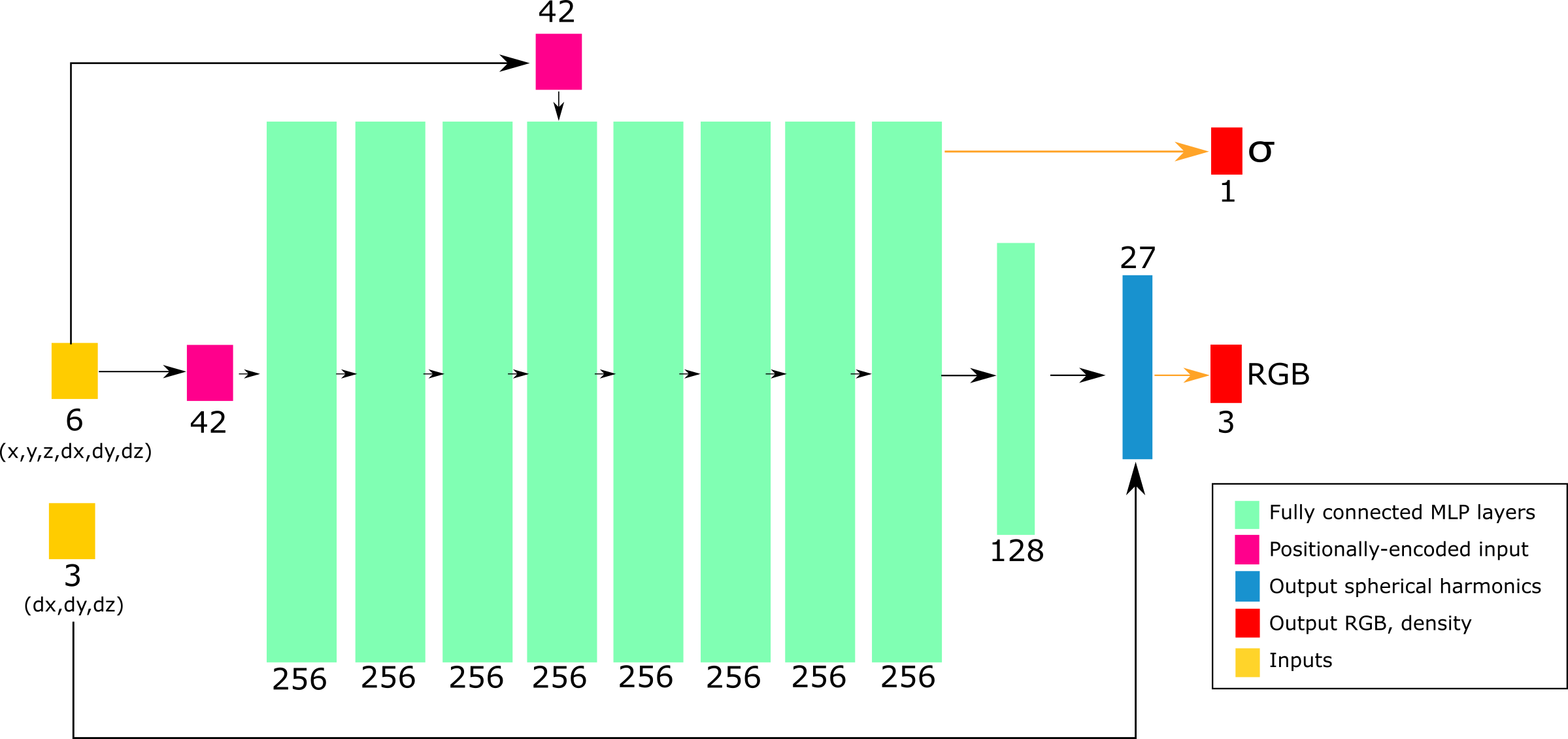}
    \caption[Model architecture]{\textbf{Our architecture.} Following previous works on NeRF, we use a densely connected 8-layered 256-neurons-wide multilayer perceptron (MLP), with an extra output layer of 128 neurons. We use ReLU as the activation function between the inner layers and a extended-range sigmoid for the final transformation from SH to RGB values.}
    \label{fig:arch}
   \end{minipage}
\end{figure}

\paragraph*{Network architecture}
\label{sc:arch}
The network takes as input the position and direction, which are augmented using Fourier-based positional encoding~\cite{mildenhall2020}, and outputs the density $\sigma(\px)$ and the $(l_{\mathrm{max}}+1)^2$ SH coefficients of the directional emission $\SH(\Phi)_{l}^{m(\px)}$ for each color channel, with $l_\mathrm{max}$ the number of SH levels.  
Following previous work~\cite{yu2021plenoctrees}, we use a fully-connected 8-layered 256-neurons-wide multilayer perceptron, with an extra output layer of 128 neurons returning $3\,(l_\mathrm{max}+1)^{2}$ coefficients of the SH expansion of $\Phi(\px, \omega)$ (Figure~\ref{fig:arch}). We use ReLU as the activation function between the inner layers.

In order to get the final emitted color $\Phi(\px,\omega)$, we also apply a non-linear activation function over the SH-projection, as
\begin{equation}
    \Phi(\px,\omega) = \mathcal{S}\left(\sum_{l=0}^{l_\mathrm{max}}\sum_{m=-l}^{l}\SH{\Phi}_l^m(\px)\,Y_l^m(\omega)\right),
\end{equation}
with $Y_l^m(\omega)$ the SH basis projected in direction $\omega$, and $\mathcal{S}(\cdot)$ an extended-range sigmoid (a sigmoid multiplied by the maximum scene radiance). This allows us to compute losses in a bounded [0,1] range. This proved much more stable than using unbounded activation functions.

\subsection{Loss function}
\label{ch:appearance:losses}
Following the original work of NeRFs~\cite{mildenhall2020}, we use two different models for supervising the learning of both the density and emission, by comparing against the training HDR image. In addition, we incorporate a term for actively supervising the transparency $\alpha(\Ray)$. Therefore, our loss function is
\begin{equation}
\mathcal{L} = \frac{1}{|\mathcal{R}|}\sum_{\Ray\in{}\mathcal{R}}\left [ \mathcal{L}_c(\Ray) + \mathcal{L}_f(\Ray) + \mathcal{L}_\alpha(\Ray) \right ],
    \label{eq:finaloss}
\end{equation}
where $\mathcal{R}$ is the set of $|\mathcal{R}|$ sampled rays,  $\mathcal{L}_c(\Ray)$ and $\mathcal{L}_f(\Ray)$ are the HDR color loss for the coarse and fine models, respectively, and $\mathcal{L}_\alpha(\Ray)$ is the transparency loss.

\paragraph*{HDR color loss} Using standard mean squared error (MSE) in high-dynamic range data results in severe problems in areas with low radiance, since MSE is dominated by errors in brighter regions (e.g. directly visible light emitters). In order to achieve an even learning across the whole dynamic range, we regularize the MSE by the square of the approximate luminance of the pixel, in the spirit of Lehtinen and colleagues~\cite{noise2noise}. In essence, Lehtinen et al. proposed to use a mapping similar to Reinhard's global tonemapping operator~\cite{reinhard02}
\footnote{Reinhard's global tonemapping has the form  $M(y) = \frac{\log(\lambda\mathrm{y} + \epsilon)}{L_{\mathrm{max}}}$ and derivative $M'(y) = \frac{\lambda}{(\lambda\mathrm{y} + \epsilon)^{-1}}$.}. However, unlike their approach, we also need to multiply the estimated pixel luminance by the maximum radiance of the luminaire $L_\mathrm{max}$ to compensate the fact that we are estimating color in the [0,1] range. We found that keeping the sigmoid activation function for color and extending its range to the maximum radiance of the luminaire enabled a more stable learning of the emission than using exponential functions to unbound the output, which by definition struggle with producing values close to the absolute 0. This results in an HDR color error loss for the coarse model (the loss for the fine model is analogous) defined as
\begin{equation}
\label{eq:tnloss}
    \mathcal{L}_c(\Ray) = \left\| \frac{\widehat{\Le^c}(\Ray) - \Le(\Ray)}{\lambda\,\widehat{\Le^c}(\Ray) + \epsilon} \right \|_{2}^{2} ,
\end{equation}
with $\widehat{\Le^c}(\Ray)$ the color prediction by the coarse network, $\Le(\Ray)$ the emission ground truth, and $\epsilon$ is a regularizing empirical term (we found that $\epsilon = 0.01$ works well for all our tested luminaires). In practice, we found that any decaying function can be used as a regularizer to the MSE loss, balancing the supervision of low and high radiance areas. For example, we also had success with exponentially-decaying regularizing functions, but these introduced an additional hyperparameter (the decay of the exponential function) that required manual tuning for each luminaire. 

\paragraph*{Transparency loss}
While the color metric above removes most problems with HDR content, we found that for scenes with isolated objects (i.e. no backgrounds) with HDR radiance, the training convergence would sometimes depend on supervising the transparency. Most importantly, achieving perfect predictions of the transparency of each ray is \new{necessary for a successful integration of the model} within a traditional rendering engine later on, enabling perfect blending of the luminaires with the rest of the scene.
We therefore introduce a direct supervision of alpha masks through a MSE loss, as
\begin{equation}
\label{eq:alphaloss}
\mathcal{L}_\alpha(\Ray) = \left \| \widehat{\alpha}^f(\Ray) - \alpha(\Ray)\right \|_{2}^{2},
\end{equation}
with $\widehat{\alpha}^f(\Ray)$ the transparency estimated by the fine network. Note that we only need to supervise opacity in the fine model, as it is the one that will define the final opacity of a ray. Since all our data is synthetic we have ground truth alpha masks available for supervision.

\subsection{Training Details}
\label{sc:train}
We used JAX \cite{jax2018github} and its machine learning API, FLAX \cite{flax2020github} for training. We trained our models with a batch size of 1024 rays, each of them sampled at $N_c = 64$ positions in the coarse model, and $N_f = 128$ additional coordinates in the fine one. We set $l_\textrm{max} = 2$ levels of spherical harmonics. We use a 100 images as our training set, and a further 100 images for testing. All three components of the loss (coarse color supervision, fine color supervision, and alpha supervision) were given the same weight. We used Adam~\cite{adam} combined with an exponentially decaying learning rate, starting at $lr = 5\times10^{-4}$ and decaying to $lr = 5\times10^{-6}$ by the end of training. Our models were trained for 700k-1.1M iterations, which took between 1 and 2 days on an Nvidia RTX 2080Ti. 


\section{Implementation details}
\label{ch:mitsuba}

We integrate our RF-based complex luminaires in  Mitsuba~\cite{jakob2010mitsuba}, as a new Emitter plugin attached to the proxy geometry (a sphere or a box) returning the emission $\Le(\Ray)$ (Equation~\eqref{eq:nerfequation}). We also implement a null-like BSDF that returns the transparency $\alpha(\Ray)$ (Equation~\eqref{eq:alpha}). Both the Emitter and the null BSDF raymarch through the volumetric representation of the radiance field, querying the density $\sigma(\px)$ and directional emission $\Phi(\px,\omega)$.

As discussed above, we do not query the trained neural network. Instead, we project the density and emission values into a plenoctree~\cite{yu2021plenoctrees}, i.e. an octree-based voxelized representation of the radiance field. This results into massive savings in terms of performance, as well as a much simpler integration in Mitsuba. Querying the octree is much cheaper than querying a neural network, and we only need to query the few intersected voxels by each ray until a surface is hit (opacity saturates), massively reducing the number of samples placed over empty space which significantly impact the performance of NeRFs, in which a set number of samples for each ray is used.

\paragraph*{Creating the plenoctree} 
We extract the octree from the neural volume by densely sampling the density $\sigma(\px)$ and the SH coefficients of the emission $\SH{\Phi_l^m}(\px)$. We first place a single sample in the center of each voxel to obtain their directionally-invariant density, and then prune voxels whose density is below an empirically set threshold (0.01). This allows us to prune empty space and most NeRF artifacts. Then, for each dense voxel, we throw an additional 256 positional samples uniformly placed within the volume defined by the voxel and average them to avoid aliasing at discontinuities.
The maximum depth of the octree directly controls the quality of the extracted representation, with higher depths enabling finer detail, while also requiring more storage space. \new{An octree using a maximum depth of 10 (fine quality), containing the representation of a single luminaire, has a memory footprint of around 1.7GBs.} 
%


\paragraph*{Sampling} In our current implementation we uniformly sample the proxy geometry, which is admittedly suboptimal and does not leverages the precomputed data of the luminaires radiance field. Importance sampling according to the RF would provide a significant variance reduction, although it remains an avenue of future work.

\begin{figure*}[t]
   \begin{minipage}[b]{\textwidth}
    \includegraphics[width=0.19\textwidth]{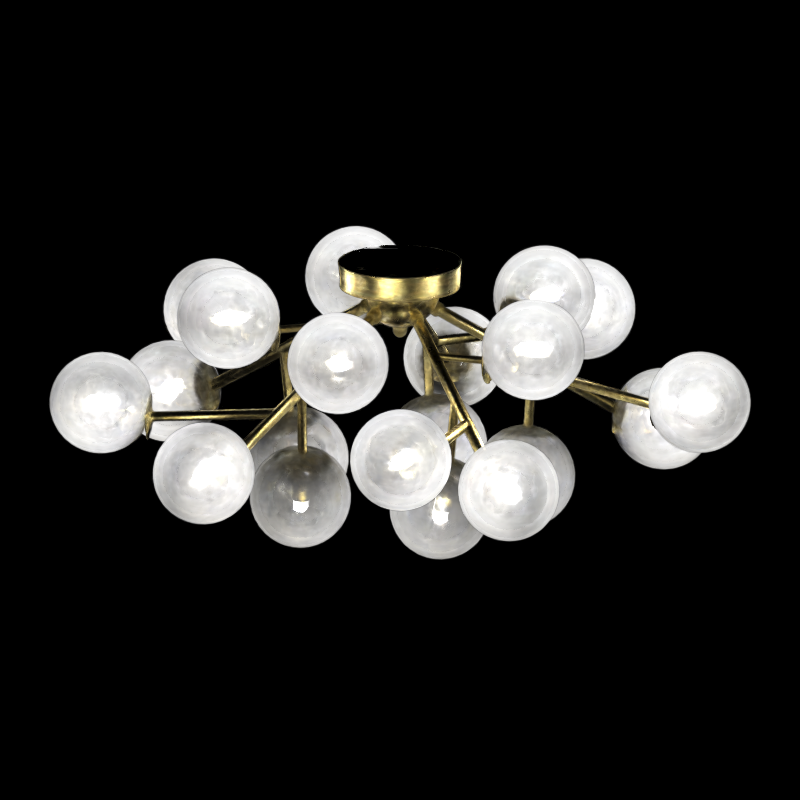}
    \includegraphics[width=0.19\textwidth]{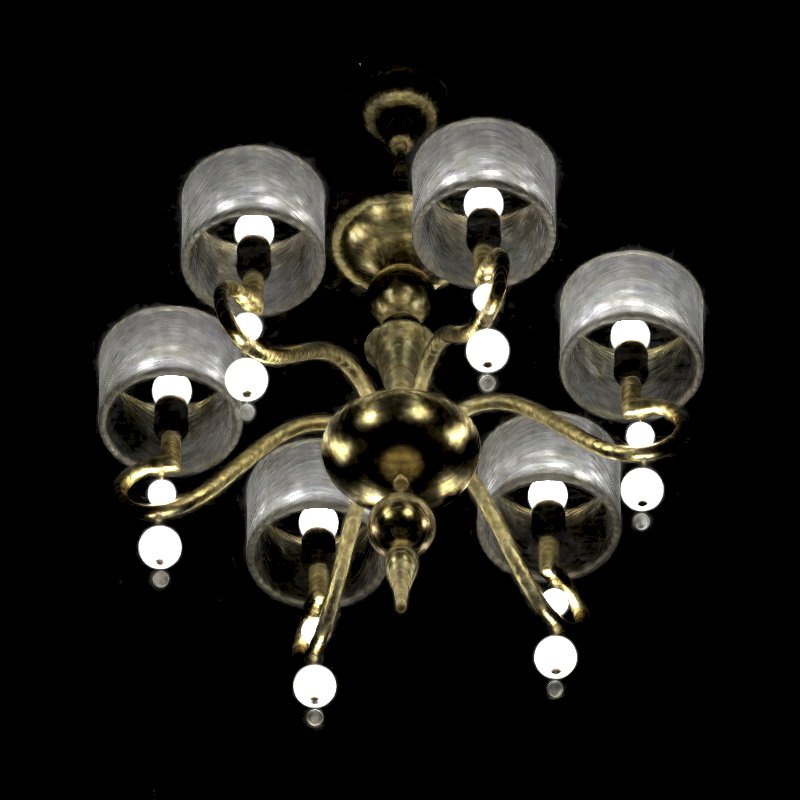}
    \includegraphics[width=0.19\textwidth]{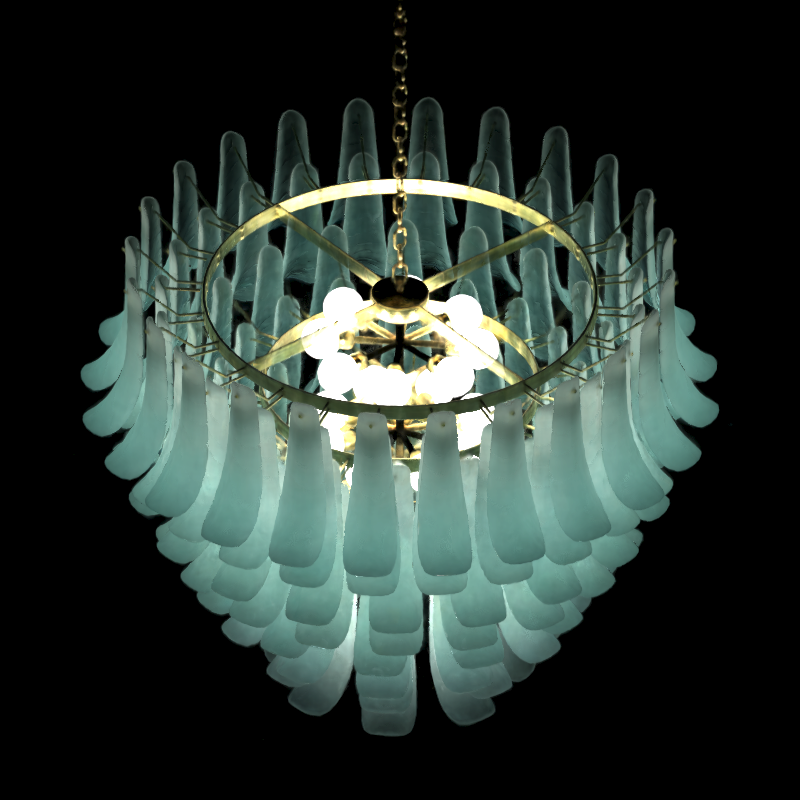}
    \includegraphics[width=0.19\textwidth]{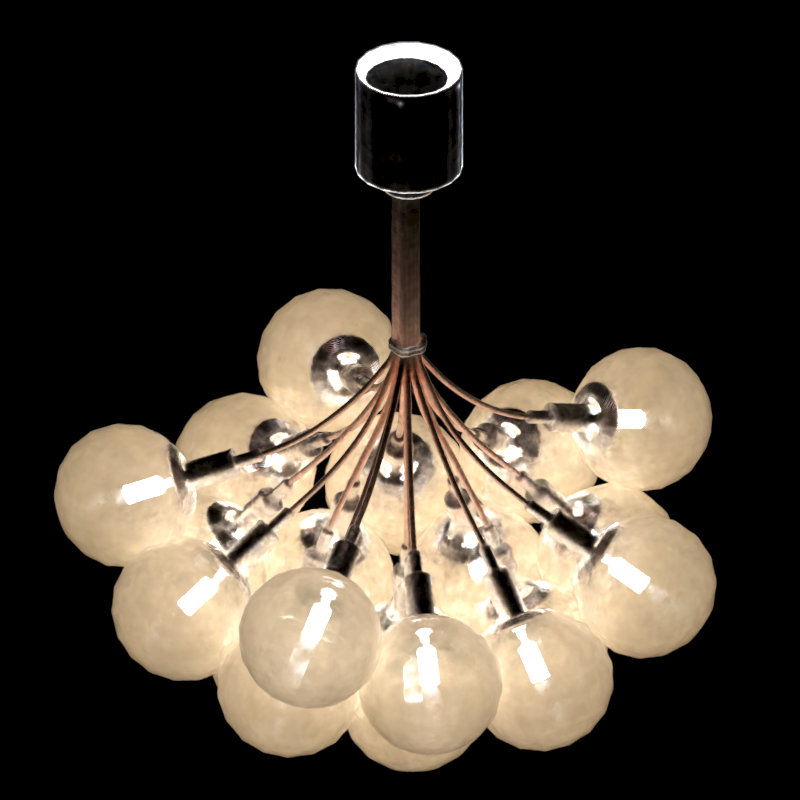}
    \includegraphics[width=0.19\textwidth]{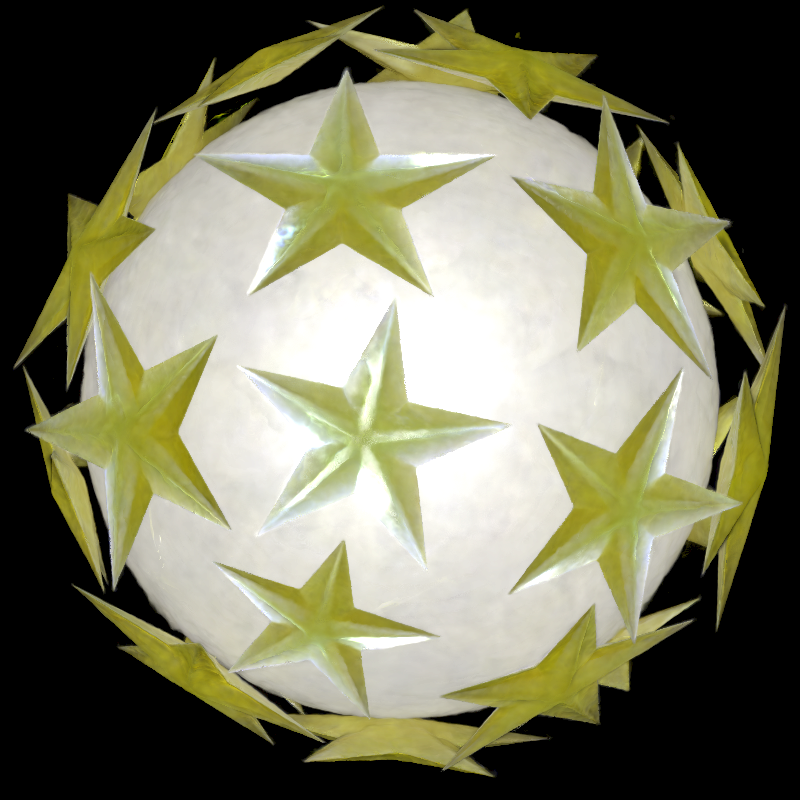}
   \end{minipage}
   \caption{\textbf{Reconstruction examples. }Example test views of the reconstruction of our dataset using our models. From left to right: \textsc{Dallas}, \textsc{Neoclassical}, \textsc{Portica}, \textsc{Cluster}, and \textsc{Star}. Average reconstruction errors for all datasets can be found in Table~\ref{table:metrics}. }
   \label{fig:testviews}
\end{figure*}

\section{Evaluation and Results}
\label{ch:results}

\subsection{Datasets}
\label{sc:datasets}
We generate our datasets in Mitsuba using a volumetric path tracer. Each luminaire is rendered using an orthographic camera placed at a 100 pseudo-random directions, generated using a 2D Halton sequence on the sphere around the luminaire. Each view is rendered with a resolution of $800^2$, making a total of 64M rays per luminaire. 
%
%

We render five different luminaires with varying geometric complexity, number of emitters, and dynamic ranges (with maximum radiance ranging from  4.5 in \Dataset{Star} to 40 in \Dataset{Cluster}). \new{We intended to create a datasets of luminaires with a size and variety similar to Zhu et al.'s \cite{zhu2021neural}, with some luminaires showing caustic paths (e.g. the light bulbs in both \textsc{Dallas} and \textsc{Cluster}), as well as diffuse appearance due to multiple scattering. The rendering time of each dataset depends on its complexity, but they took between 12 and 36h each using 32 threads in a workstation with a dual Intel Xeon Gold processor.} Unless stated otherwise, our results are generated using an order-2 SH expansion ($l_\mathrm{max}=2$, 9 SH coefficients), with a maximum octree depth of 10 levels (spatial resolution of $1024^3$).


\subsection{Evaluation}
Here we analyze the reconstruction error introduced by our RF approximation of the luminaires, as well as the impact of our design choices on the final error. In all cases, we report the average PSNR and SSIM for all test views. 

Figure~\ref{fig:testviews} shows example reconstructions for all our datasets, with the average reconstruction errors listed in Table~\ref{table:metrics}. We refer to the supplemental video for animations of the luminaires, to assess temporal stability of our reconstructions. 



\begin{table}[h]
\centering
\begin{tabular}{lll}
\cline{2-3}
\multicolumn{1}{l|}{}                                          & \multicolumn{1}{l|}{PSNR}             & \multicolumn{1}{l|}{SSIM}            \\ \hline
\multicolumn{1}{|l|}{\Dataset{Dallas}}                      & \multicolumn{1}{l|}{31.5072}          & \multicolumn{1}{l|}{0.9374}          \\ \hline
\multicolumn{1}{|l|}{\Dataset{Portica}}                                      & \multicolumn{1}{l|}{28.6694}      & \multicolumn{1}{l|}{0.8594}     \\ \hline
\multicolumn{1}{|l|}{\Dataset{Cluster}}                          & \multicolumn{1}{l|}{27.7260}          & \multicolumn{1}{l|}{0.8872}          \\ \hline
\multicolumn{1}{|l|}{\Dataset{Neoclassical}}              & \multicolumn{1}{l|}{26.8784}          & \multicolumn{1}{l|}{0.8829}          \\ \hline
\multicolumn{1}{|l|}{\Dataset{Star}}              & \multicolumn{1}{l|}{27.8027}          & \multicolumn{1}{l|}{0.9119}          \\ \hline
\end{tabular}
\caption[Results Metrics]{\textbf{Reconstruction error. } PSNR and SSIM error metrics for all datasets used in our work, averaged over all test views of the luminaires.}
\label{table:metrics}
\end{table}

\begin{figure}[!htbp]
   \begin{minipage}[b]{\columnwidth}
    \includegraphics[width=0.31\columnwidth]{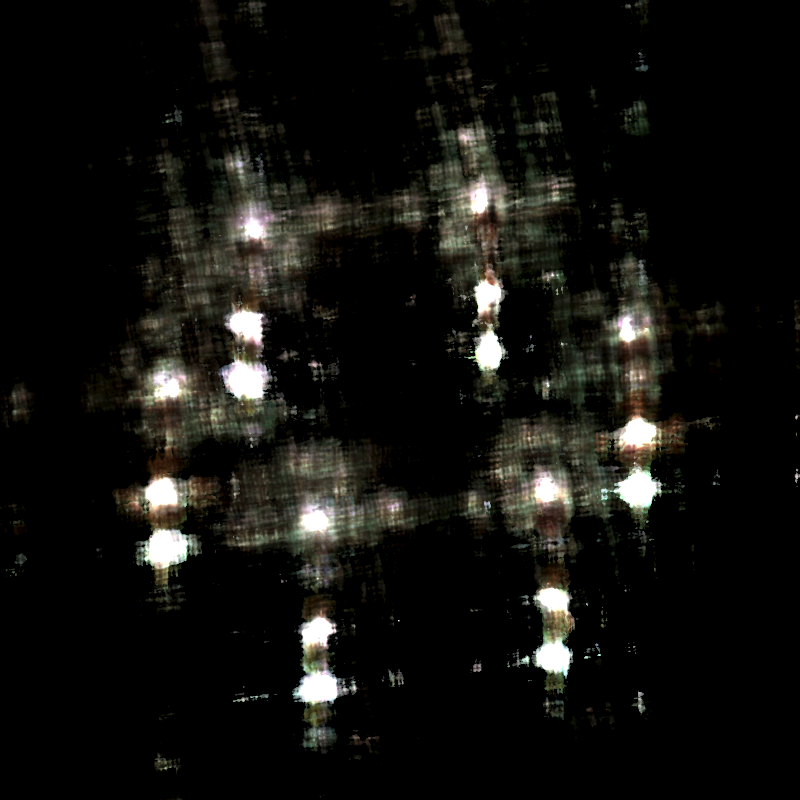}
    \includegraphics[width=0.31\columnwidth]{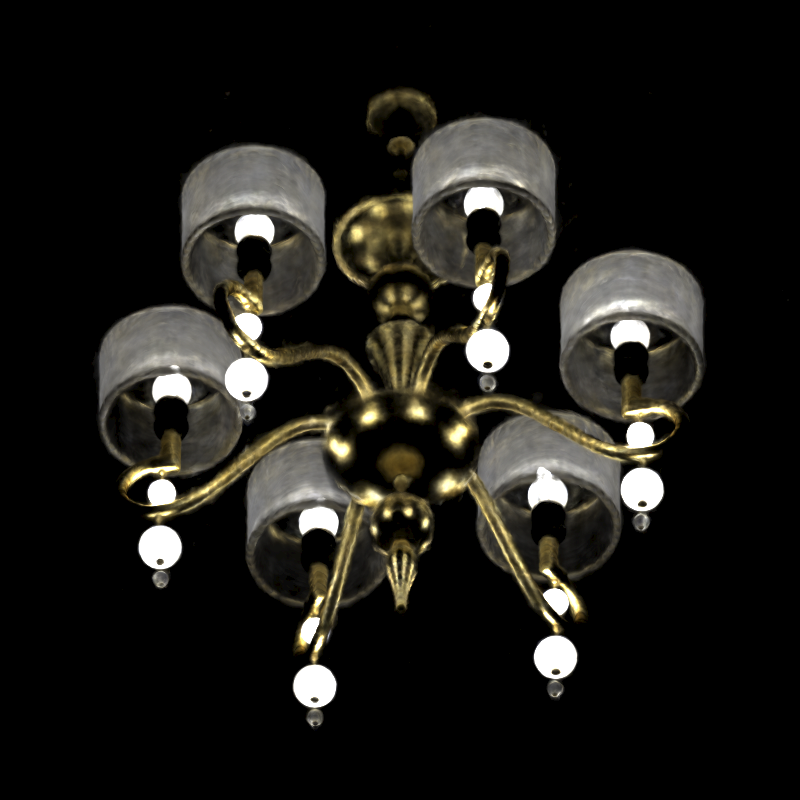}
    \includegraphics[width=0.31\columnwidth]{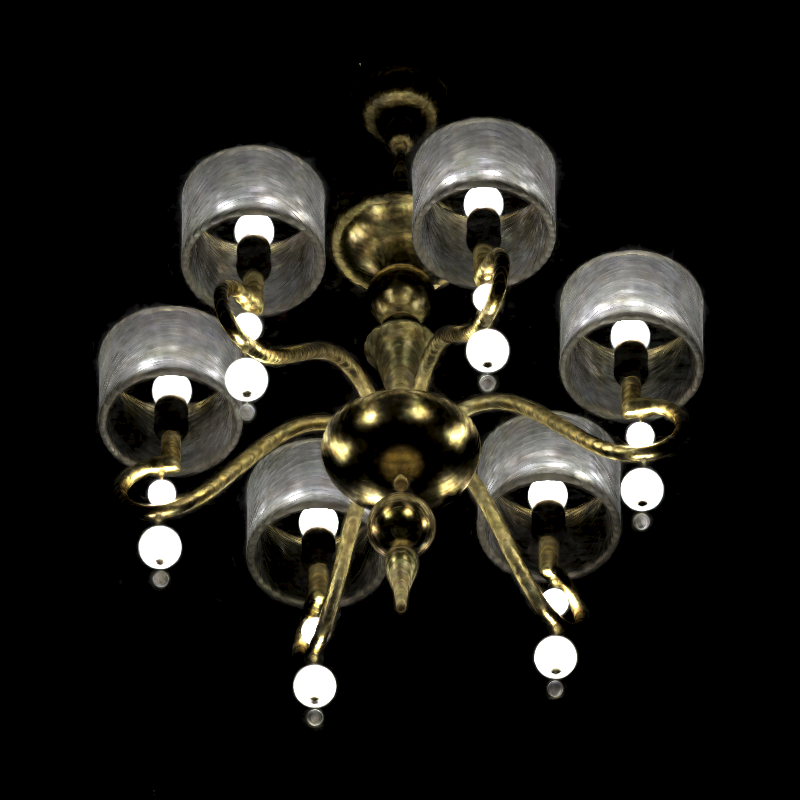}
   \end{minipage}
   \begin{minipage}[b]{\columnwidth}
    \includegraphics[width=0.31\columnwidth]{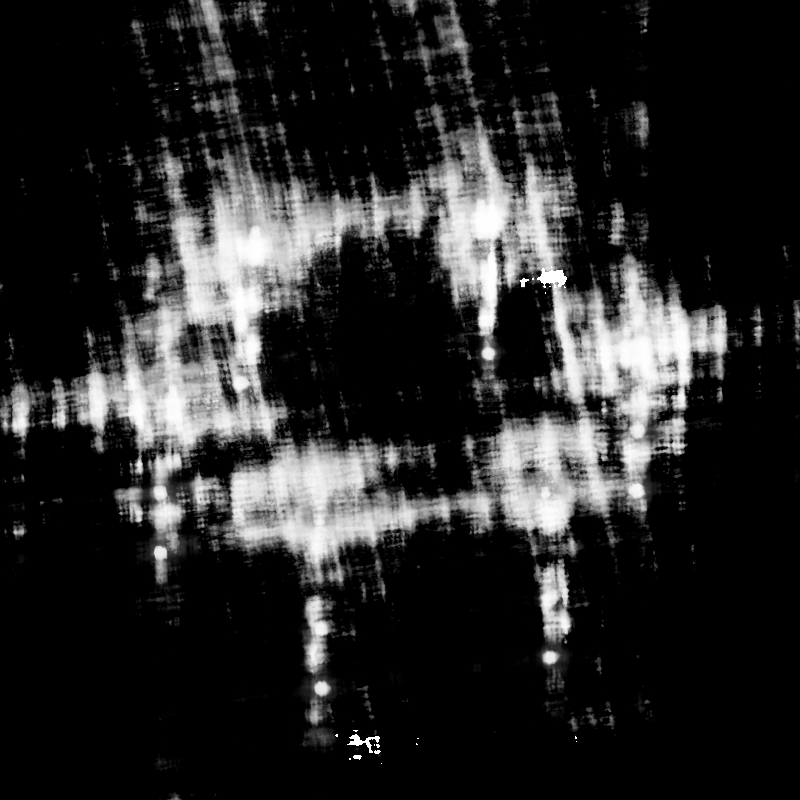}
    \includegraphics[width=0.31\columnwidth]{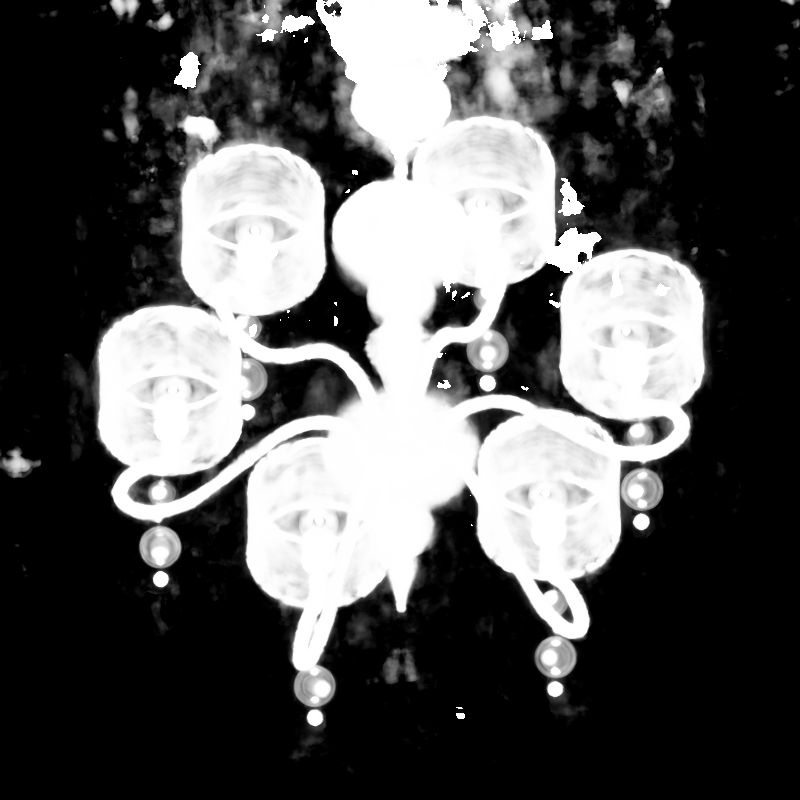}
    \includegraphics[width=0.31\columnwidth]{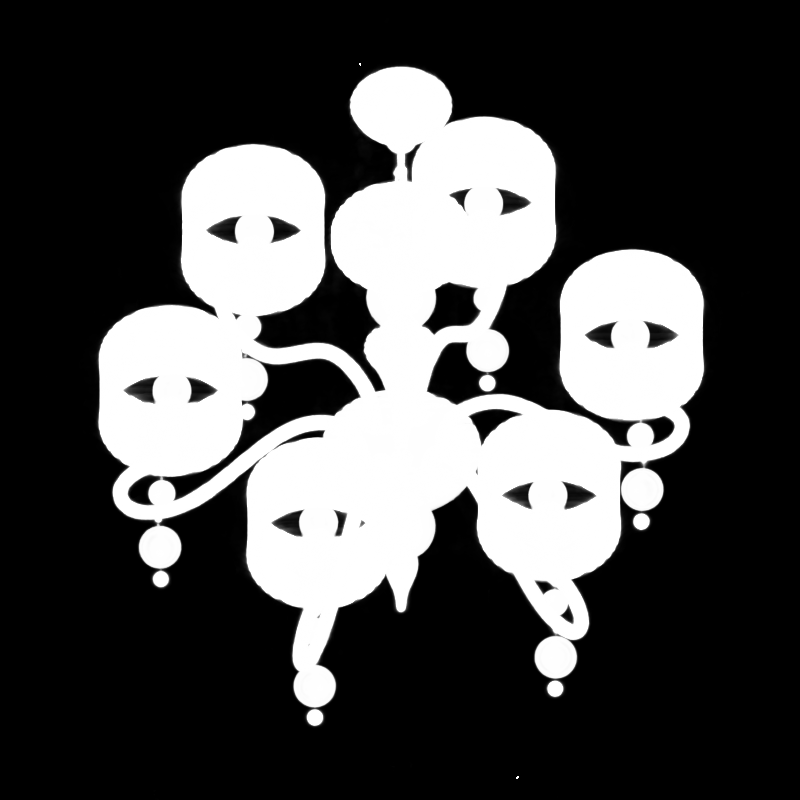}
   \end{minipage}
   \caption[Loss Ablation Study]{\textbf{Loss ablation study. } We analyze the effect of our metric on the \Dataset{Neoclassical} luminaire, in the reconstruction of both the emission $\Le$ (top) and transparency $\alpha$ (bottom). From left to right: LDR MSE, MSE+regularization,  MSE+regularization+opacity loss. The error metrics can be found in Table~\ref{table:ablation}.}
   \label{fig:ablation}
\end{figure}

\paragraph*{Loss ablation study} We analyze the effect of the different terms of our loss function \new{(HDR regularizer, transparency supervision)} in Figure \ref{fig:ablation}, for the \Dataset{Neoclassical} luminaire. \new{We compare them against the baseline NeRF loss, which uses MSE.} Our custom loss enables stable, high quality learning of the full dynamic range, and a correct volume density representation. Table~\ref{table:ablation} includes the numerical error for this experiment.
While we found that for some luminaires the transparency supervision results in a slightly worse emission reconstruction, the improvement on transparency is significant. Note that a good transparency estimation is \new{crucial} for high-quality integration of the luminaire in synthetic scenes. 


\begin{table}[!htbp]
\centering
\begin{tabular}{l|l|l|l|}
\cline{2-4}
                                                      & PSNR  $\Le$      & SSIM $\Le$       & RMSE $\alpha$ \\ \hline
\multicolumn{1}{|l|}{MSE}                             &  19.8547    &  0.6987     &    0.3617          \\ \hline
\multicolumn{1}{|l|}{MSE+reg}              &  \textbf{28.1237}    &  \textbf{0.9062}     &    0.1853          \\ \hline
\multicolumn{1}{|l|}{MSE+reg+$\alpha$ loss} &  26.8784    &  0.8829     &    \textbf{0.0358}          \\ \hline
\end{tabular}
\caption[Loss Ablation Study: Metrics]{\textbf{Loss ablation study. } Error metrics demonstrating the effect of each term in our metric for the \Dataset{Neoclassical} dataset. Note that the opacity loss (\emph{$\alpha$ loss}) slightly introduces error on the reconstruction of the luminaire emission $\Le$, but dramatically improves the reconstruction of the transparency $\alpha$.}
\label{table:ablation}
\end{table}

\paragraph*{Performance vs Quality} 
In Table \ref{table:ablationSH} we compare the performance of increasing the level of spherical harmonics when modeling the directional emission $\Phi(\px,\omega)$. The effect of increasing from $l_\mathrm{max}=2$ to $l_\mathrm{max}=4$ is marginal in error, but significantly increases rendering time, memory and storage requirements. \new{Training directly our model using a band-limited SH-based angular representation of $\Phi(\px,\omega)$, allows us to preserve high-frequencies on the ray domain (e.g. occlusions), despite the local $\Phi(\px,\omega)$ is low frequency. }

%
In order to further increase performance, we discard the integration of samples that have a density value below a certain threshold $\sigma_\mathrm{min}$, and consider a ray saturated when transparency reaches another threshold $\alpha_\mathrm{max}$. This allows us to reduce the computational overhead of multiplying spherical harmonic weights by the sampled tree value for every intersected voxel, and avoids unnecessarily querying the octree more times than those really needed to obtain an accurate color value. We have empirically set the values of these heuristics to $\sigma_\mathrm{min}=0.1$ and $\alpha_\mathrm{max}=0.9$ respectively. \new{Our method introduces an average overhead of 20\% per sample over explicitly handling the luminaires. }



\begin{table}[]
\centering
\begin{tabular}{lll}
\hline
\multicolumn{1}{|l|}{SH levels} & \multicolumn{1}{l|}{PSNR} & \multicolumn{1}{l|}{SSIM}  \\ \hline
\multicolumn{1}{|l|}{$l_\mathrm{max} = 2$ (9 coeff.)}     & \multicolumn{1}{l|}{27.7260}     & \multicolumn{1}{l|}{\textbf{0.8872}}             \\ \hline
\multicolumn{1}{|l|}{$l_\mathrm{max} = 3$ (16 coeff.)}    & \multicolumn{1}{l|}{27.6958}     & \multicolumn{1}{l|}{0.8860}                     \\ \hline
\multicolumn{1}{|l|}{$l_\mathrm{max} = 4$ (25 coeff.)}    & \multicolumn{1}{l|}{\textbf{27.7367}}     & \multicolumn{1}{l|}{0.8859}               \\ \hline
\end{tabular}
\caption[SH ablation]{\textbf{Effect of number of SH levels.} PSNR and SSIM error metrics as a function of the maximum SH levels used to model the directional emission $\Phi(\px,\omega)$ in the \Dataset{Cluster} dataset. }
\label{table:ablationSH}
\end{table}

\textbf{Transmittance model} We compare the impact of using a linear transmission model vs the classic exponential transmittance (Table \ref{table:linearmodel}). The linear model slightly outperforms the exponential one on every tested dataset, both in quality and rendering time. The faster decay of the linear transmittance reduces the number of queries to the octree, resulting in rendering times 5\% faster. 

\begin{table}[h]
\centering
\begin{tabular}{lcccc}
\cline{2-5}
\multicolumn{1}{c|}{}& \multicolumn{2}{|c|}{\Dataset{Portica}} & \multicolumn{2}{|c|}{\Dataset{Dallas}}
\\ \cline{2-5}
 \multicolumn{1}{c|}{}
 & \multicolumn{1}{c|}{PSNR} & \multicolumn{1}{c|}{SSIM} & \multicolumn{1}{|c|}{PSNR} & \multicolumn{1}{|c|}{SSIM}\\ \hline
\multicolumn{1}{|l|}{Exponential}         & \multicolumn{1}{|c|}{28.43737}     & \multicolumn{1}{c|}{0.8566}  & \multicolumn{1}{|c|}{31.4002}     & \multicolumn{1}{|c|}{0.9371}  \\ \hline
\multicolumn{1}{|l|}{Linear}              & \multicolumn{1}{|c|}{\textbf{28.6694}}     & \multicolumn{1}{c|}{\textbf{0.8594}}  & \multicolumn{1}{|c|}{\textbf{31.5072}}     & \multicolumn{1}{|c|}{\textbf{0.9374}}   \\ \hline

\end{tabular}

\caption[Linear Model Comparison]{\textbf{Transmittance model comparison.} Comparison between the classic exponential transmittance and the linear model used for our luminaires. We consistently found the linear model outperforms the exponential one in the context of modelling luminaires.} 
\label{table:linearmodel}
\end{table}

\subsection{Results} 
\begin{figure*}[!htbp]
   \begin{minipage}[b]{\textwidth}
    \includegraphics[width=0.33\textwidth]{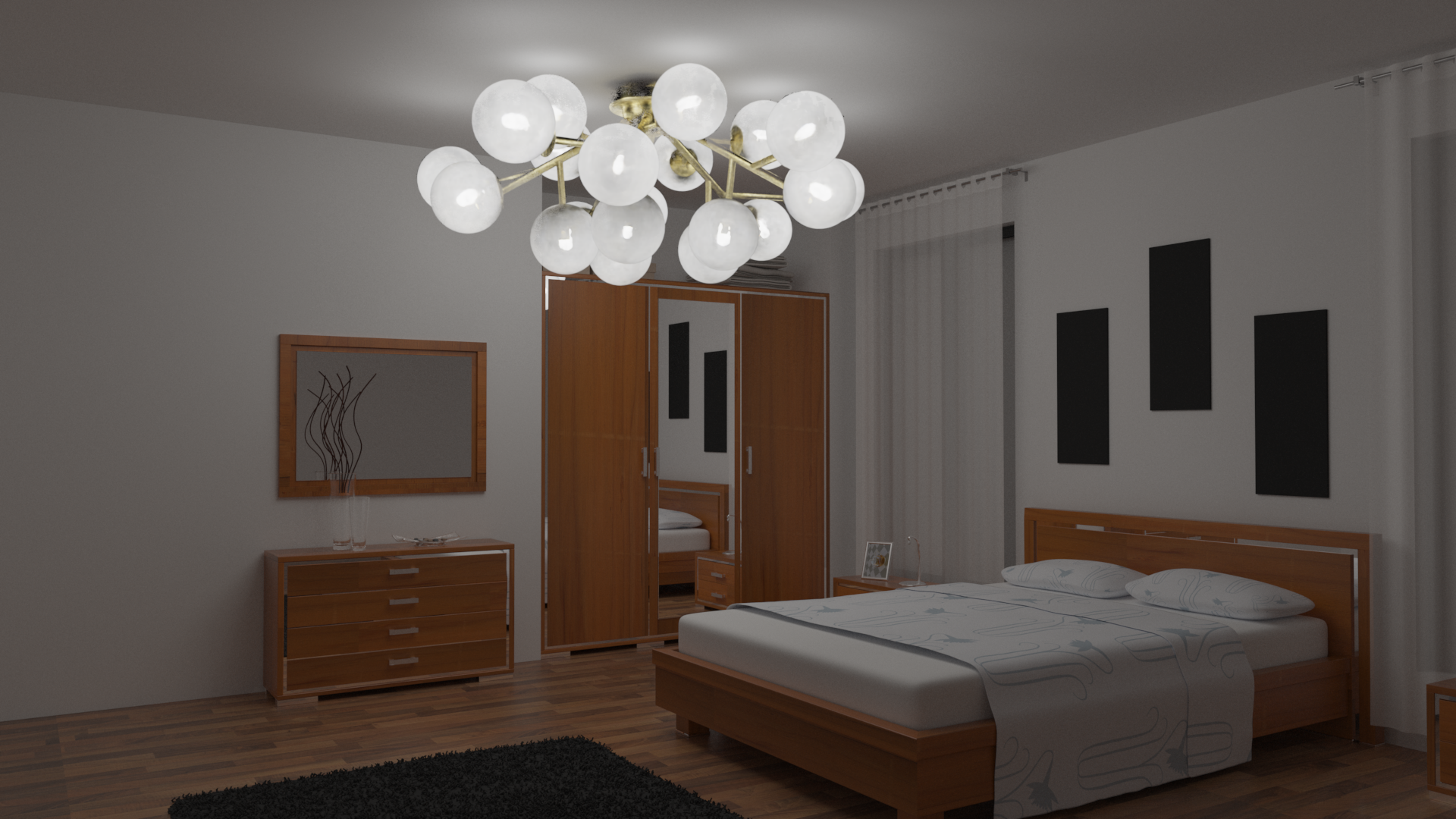}
    \includegraphics[width=0.33\textwidth]{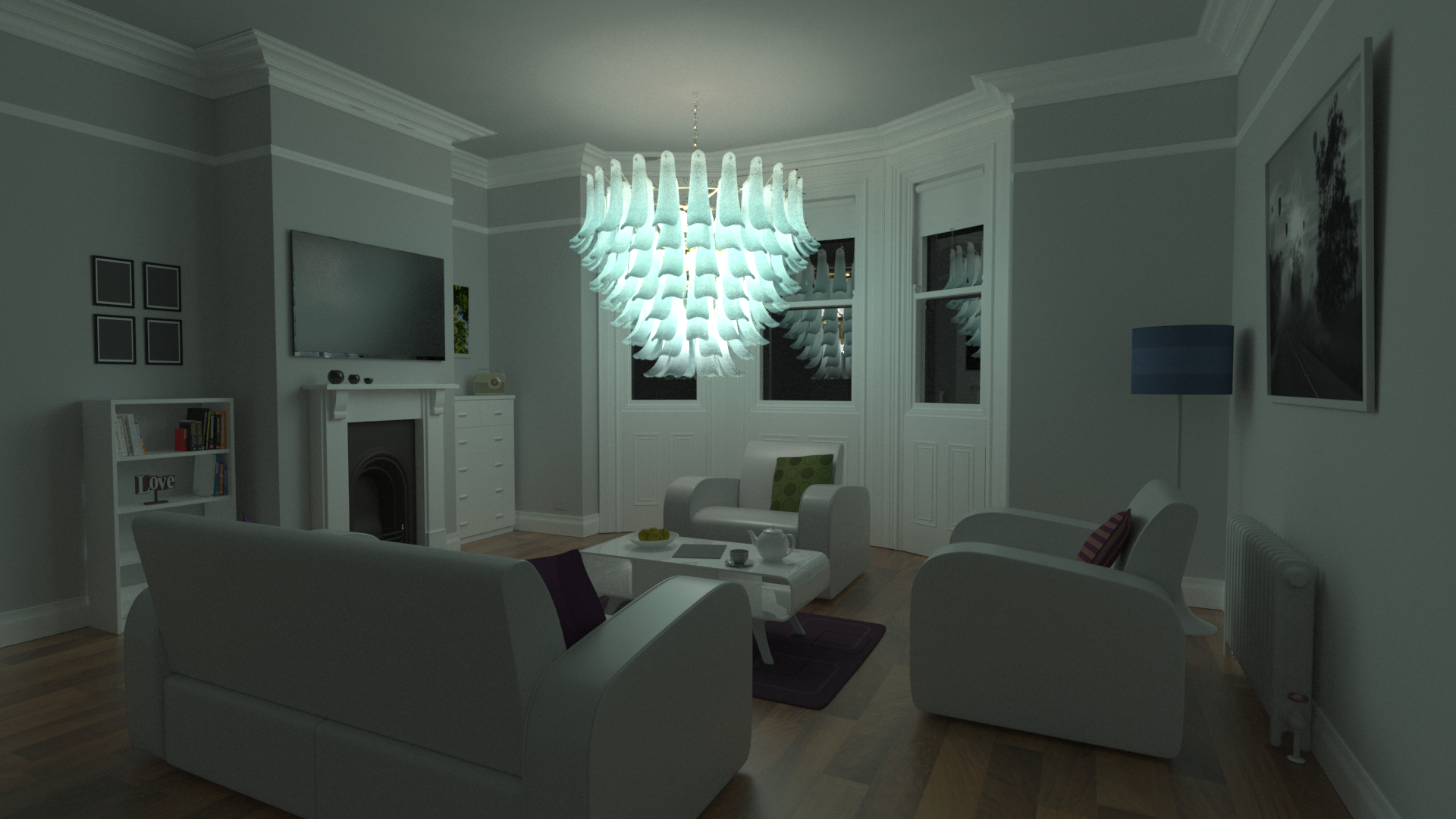}
    \includegraphics[width=0.33\textwidth]{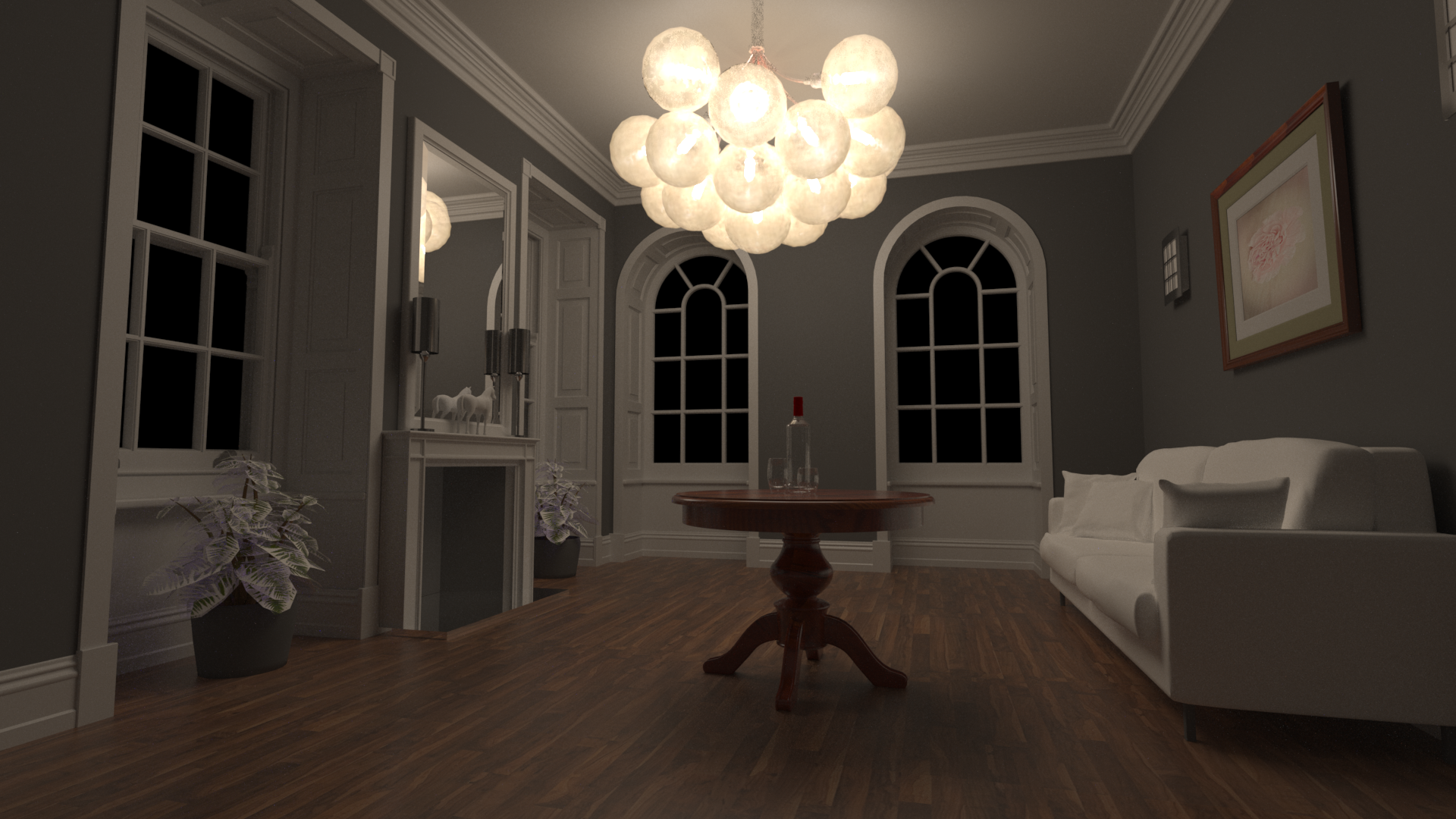}
    
   \end{minipage}
   \caption[]{\textbf{Converged results. } We render some of our scenes with more samples (4096) using our method to showcase their quality. Notice how variance is smaller in these scenes than in the reference renders used for equal error measurements, at a fraction of the rendering time.}
   \label{fig:finalrenders_highquality}
\end{figure*}

\begin{figure*}[!htbp]
   \begin{minipage}[b]{\textwidth}
    \includegraphics[width=0.24\textwidth]{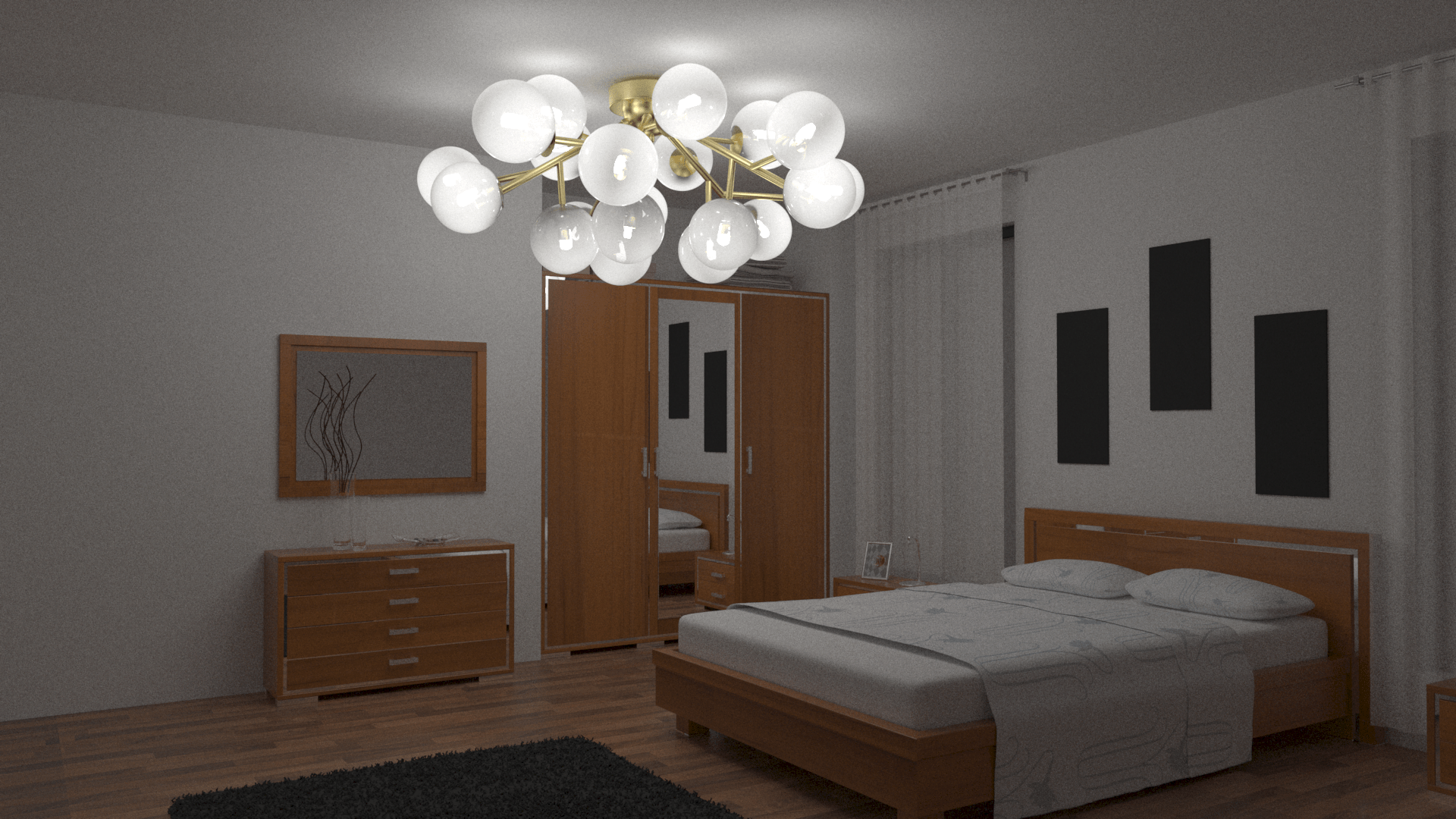}
    \includegraphics[width=0.24\textwidth]{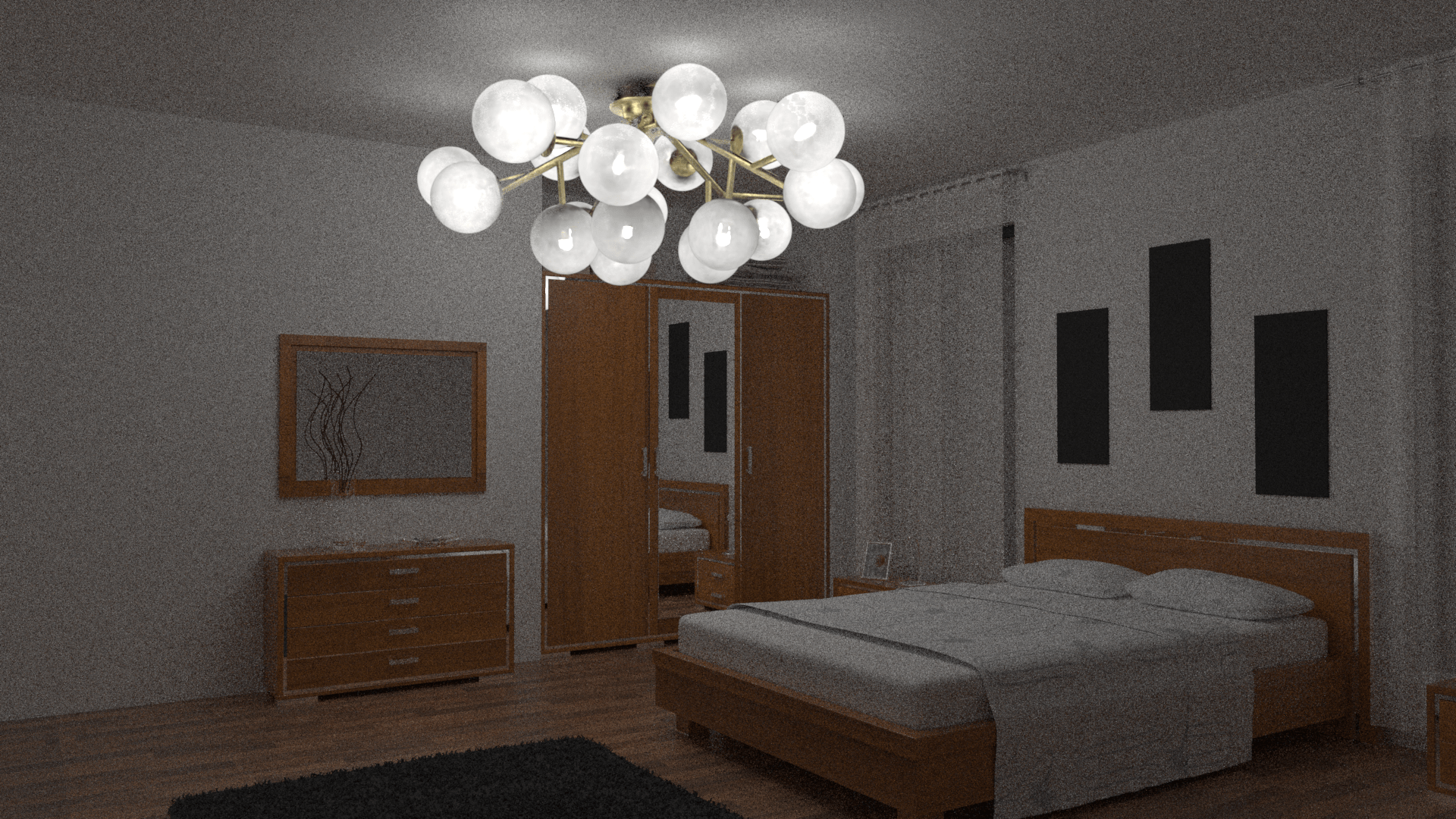}
    \includegraphics[width=0.24\textwidth]{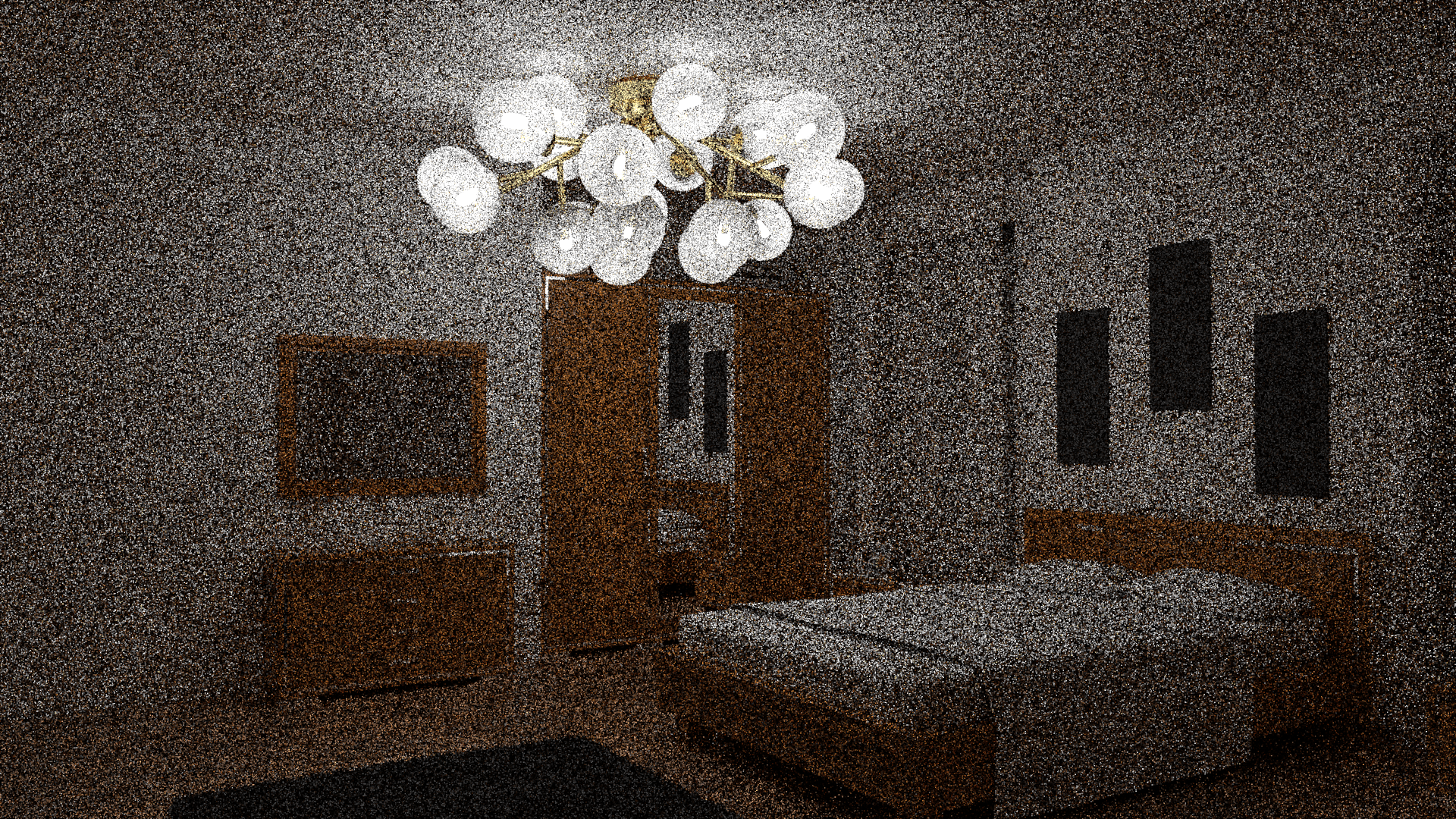}
    \includegraphics[width=0.24\textwidth]{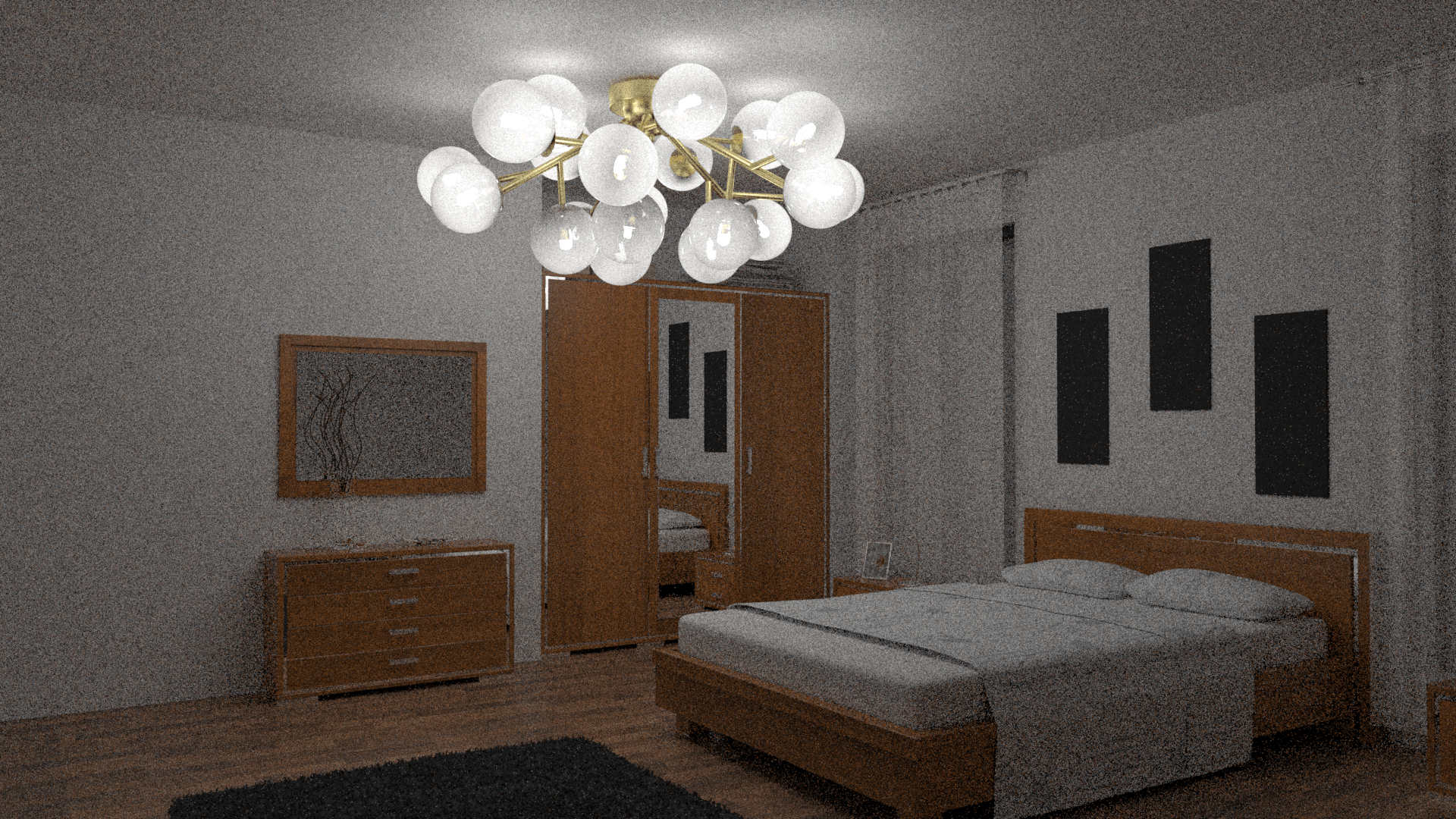}
    
   \end{minipage}
   \begin{minipage}[b]{\textwidth}
     \includegraphics[width=0.24\textwidth]{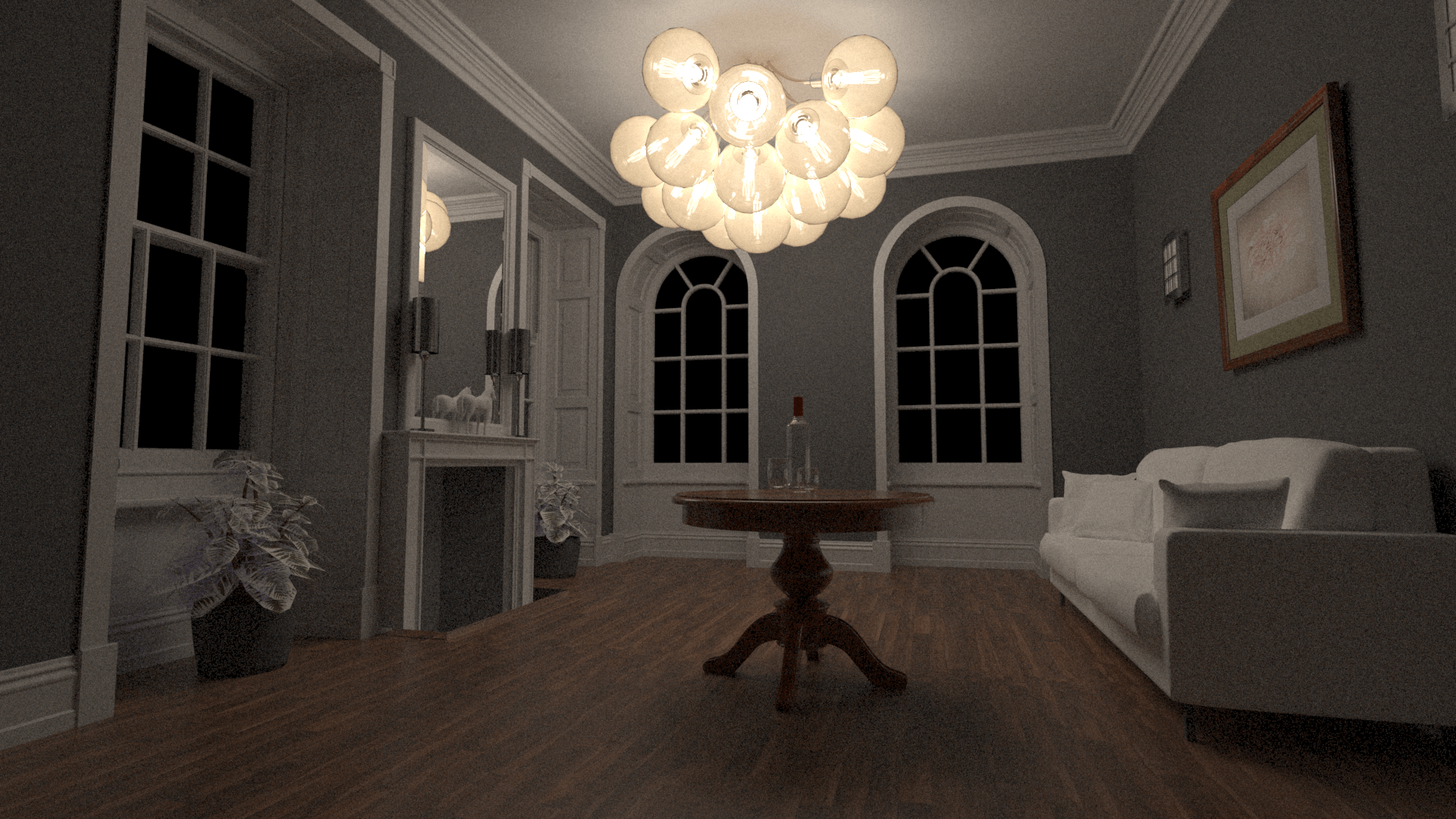}
     \includegraphics[width=0.24\textwidth]{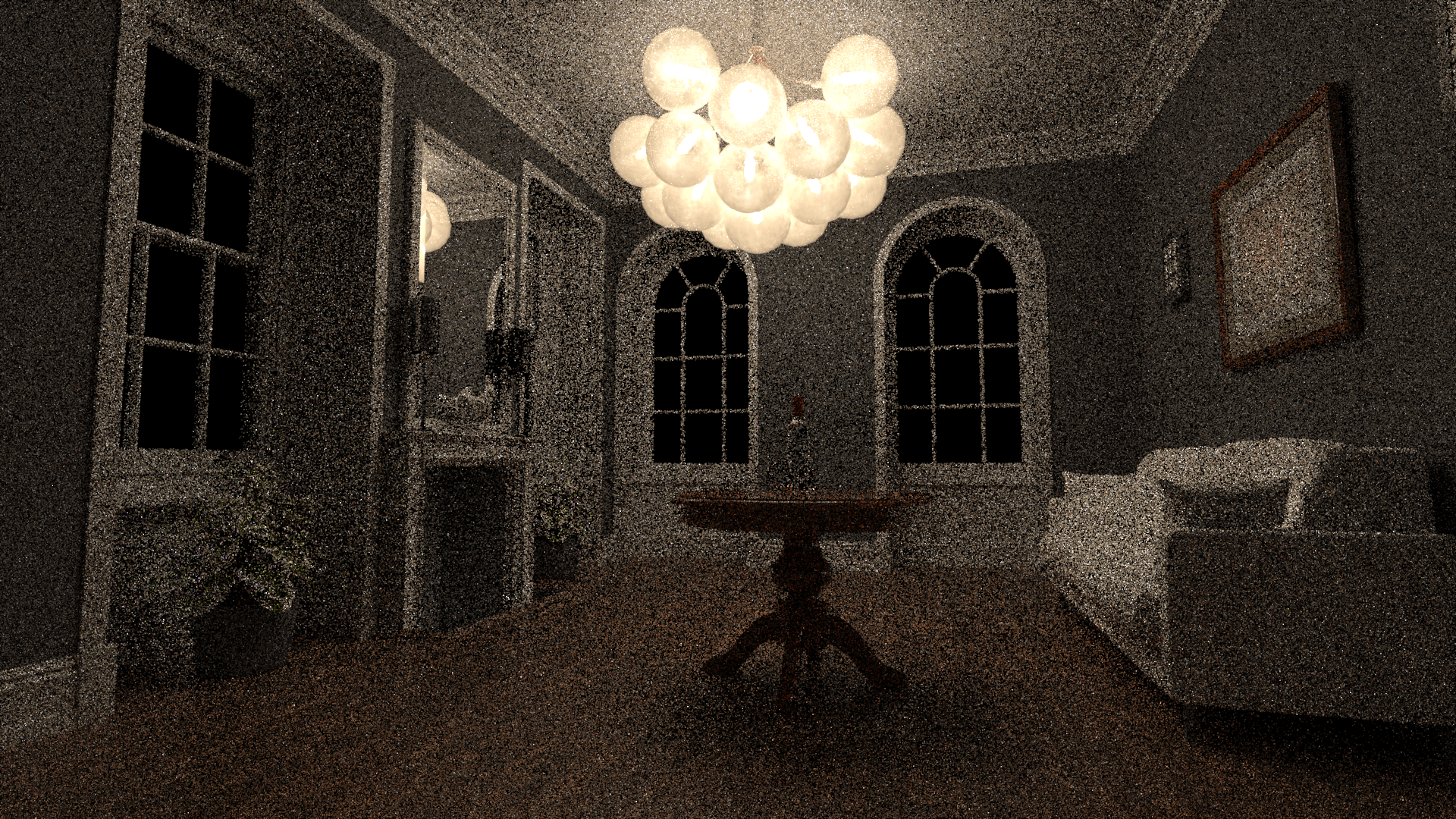}
     \includegraphics[width=0.24\textwidth]{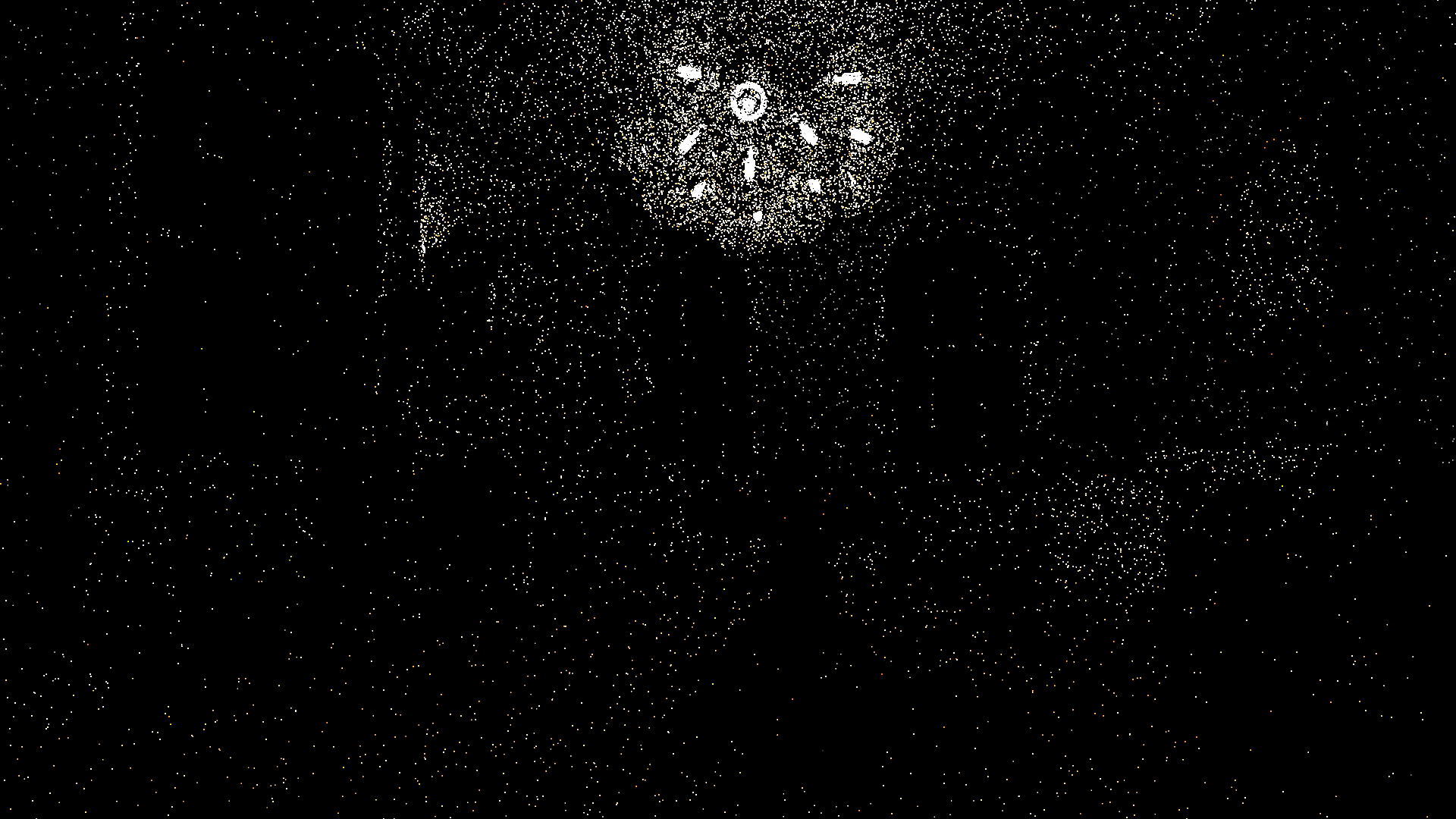}
    \includegraphics[width=0.24\textwidth]{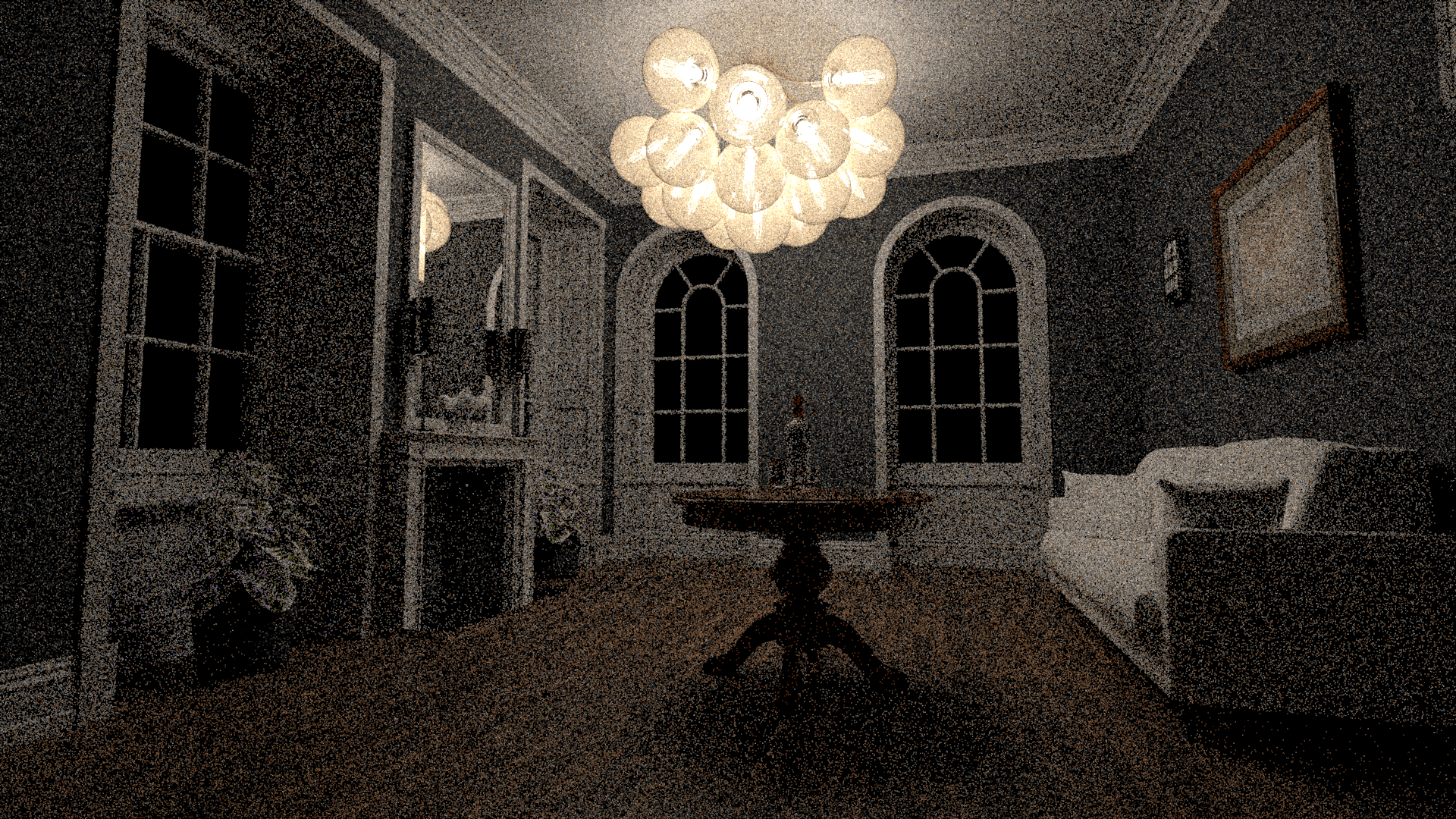}
    
   \end{minipage}
   \begin{minipage}[b]{\textwidth}
     \includegraphics[width=0.24\textwidth]{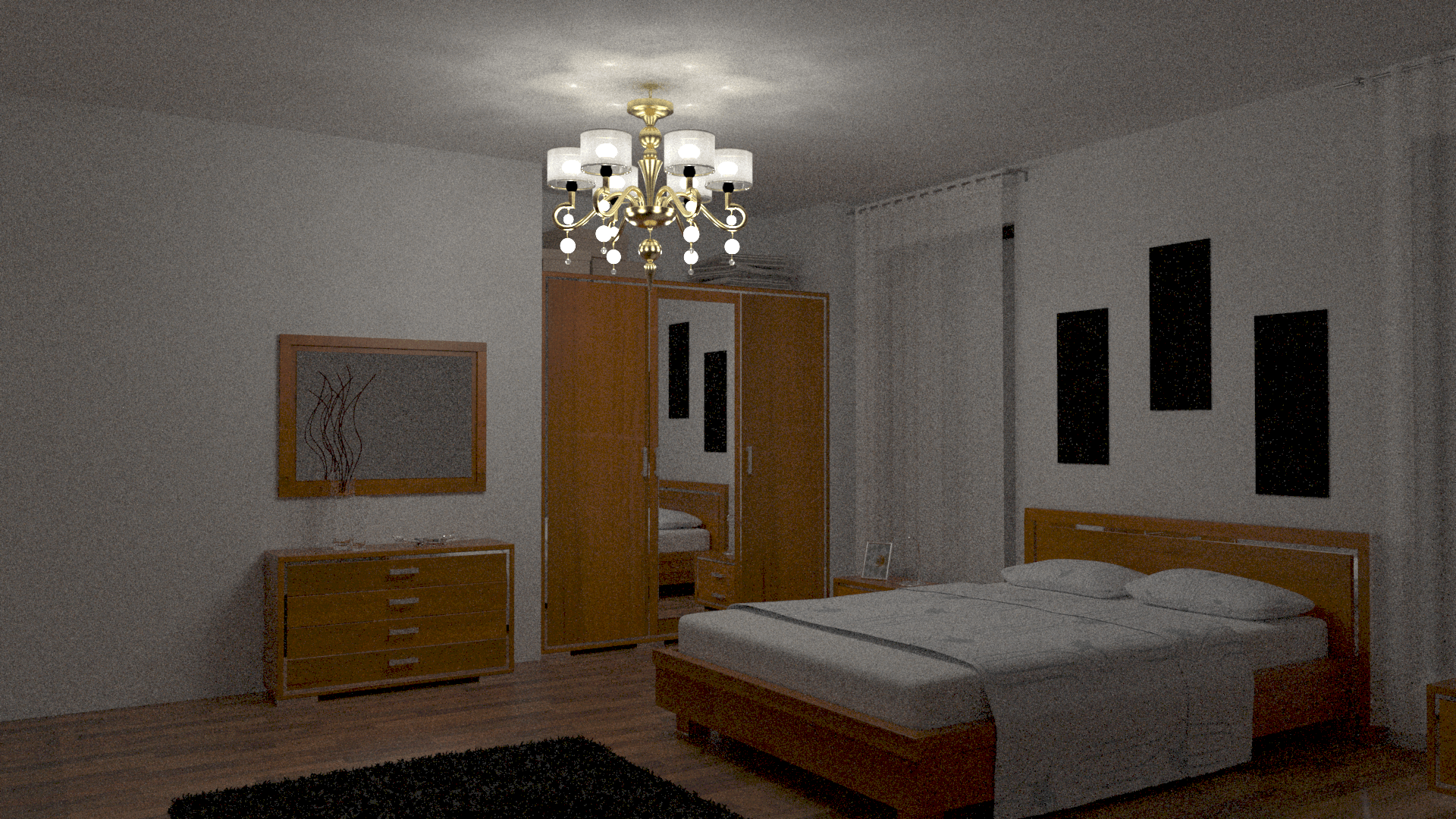}
     \includegraphics[width=0.24\textwidth]{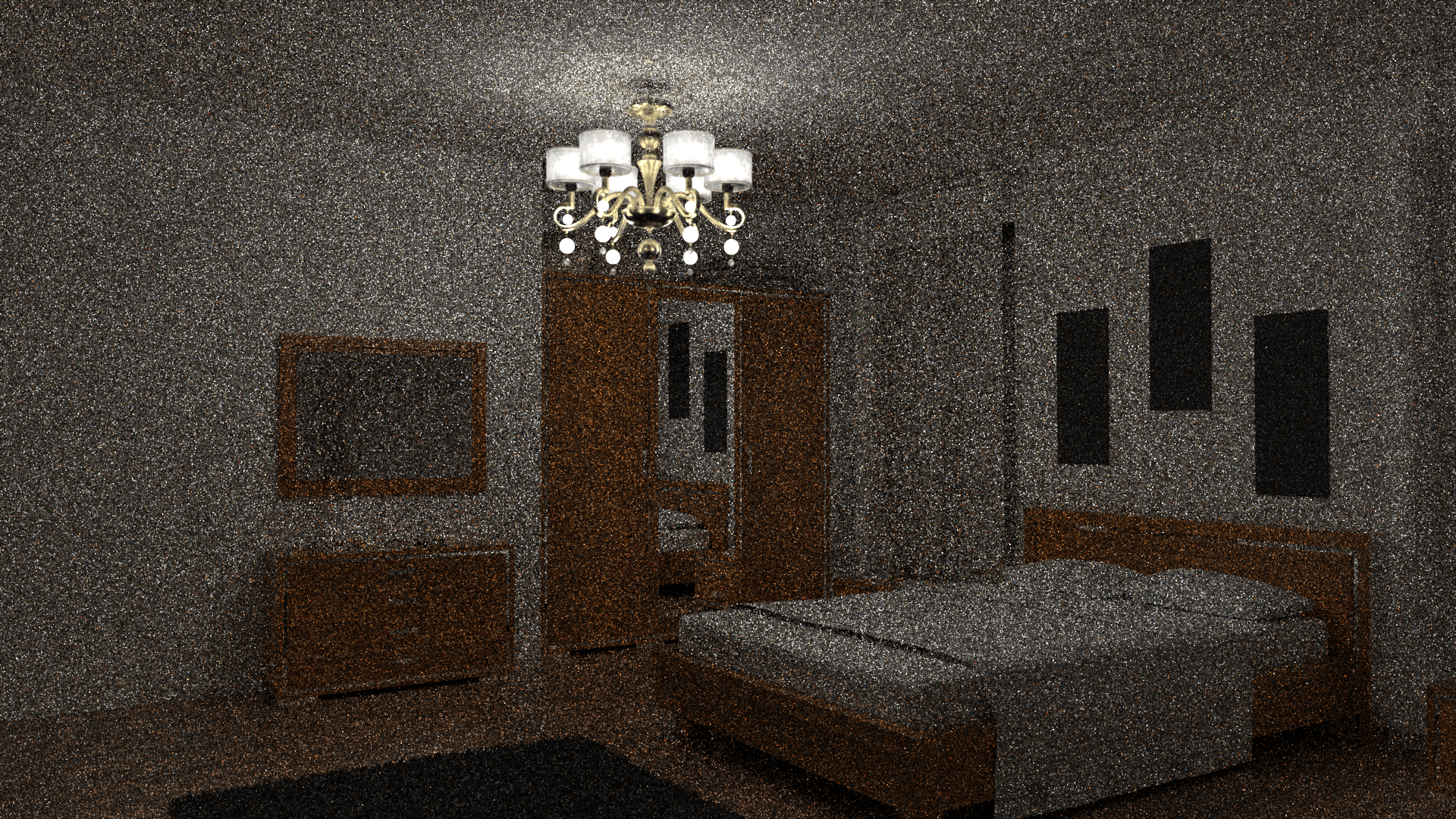}
     \includegraphics[width=0.24\textwidth]{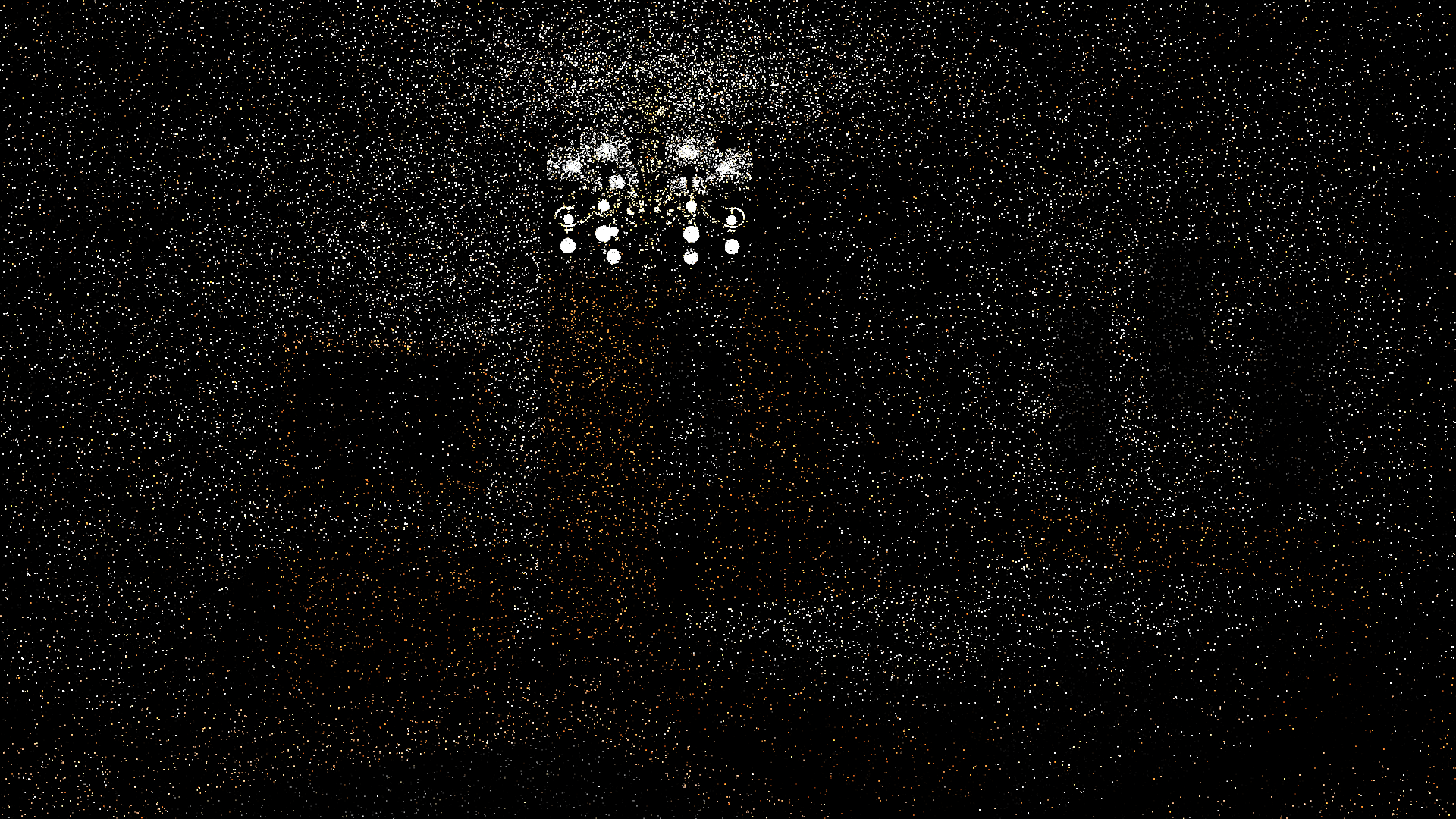}
    \includegraphics[width=0.24\textwidth]{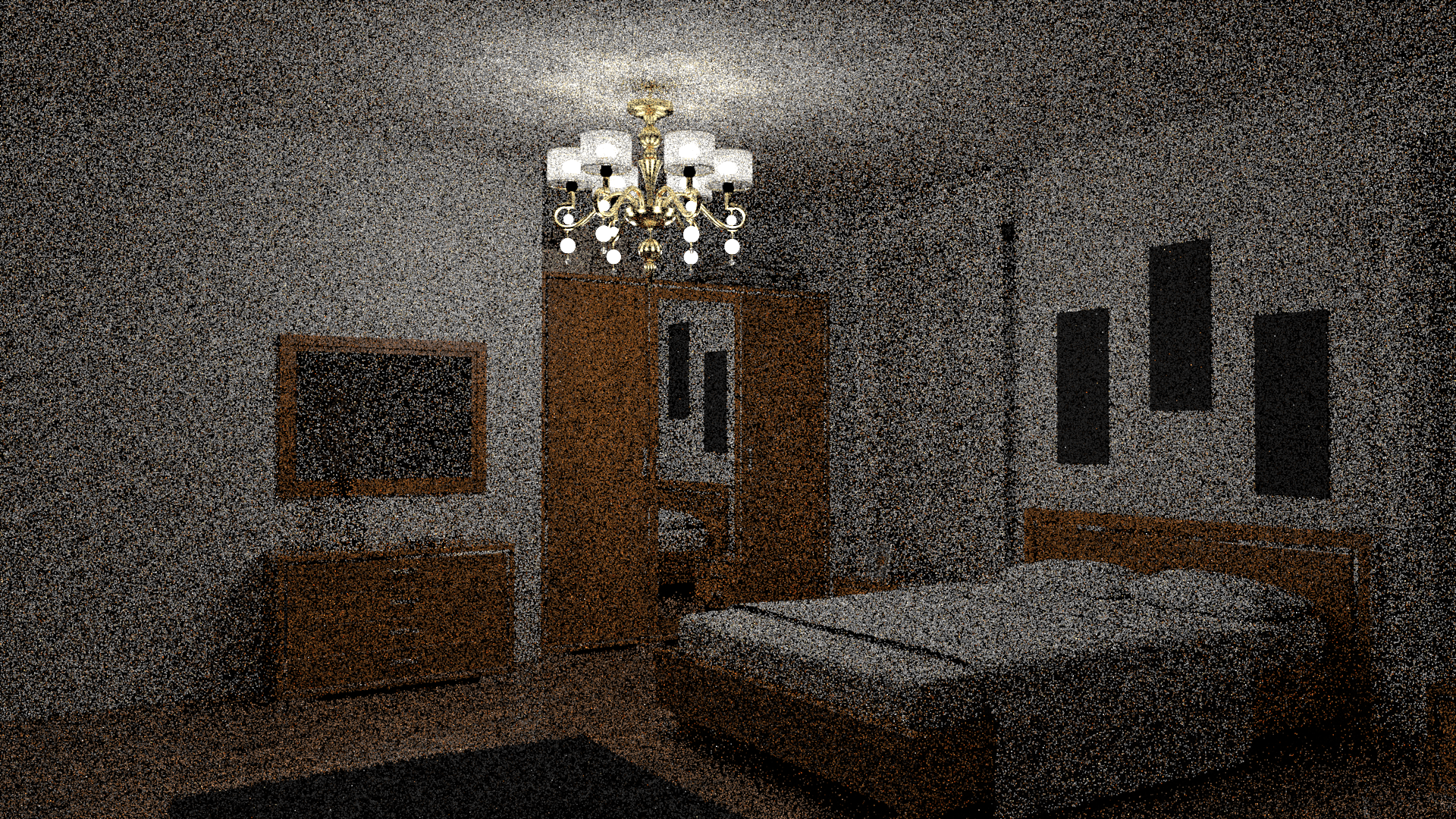}
    
   \end{minipage}
   \caption[]{\textbf{Equal-time and equal-quality comparison. }From left to right: reference render; our method (see Table \ref{table:times} for samples per pixel and rendering time), equal-time explicit renders; and equal-quality (RMSE) explicit rendering. See Table \ref{table:times} for error and rendering times. Notice how our approach is significantly faster than explicit PT or VPT while introducing minimal error in the scene.}
   \label{fig:finalrenders}
\end{figure*}

Here we show the results of integrating our RF-based volumetric representation of complex luminaires in Mitsuba. We compare our method against using the explicit light source as geometry, using path tracing (PT) or volumetric path tracing (VPT) in scenes that require the modelling of participating media (some of the tinted glasses are modelled as dielectric pieces with homogeneous mediums inside them). 
All our renders have been computed in a workstation with a dual Intel Xeon Gold processor (18x2 hyper-threaded cores), using 32 threads. Reference images are rendered with VPT or PT on the explicit luminaires representation, using 32768 samples per pixel.
%
%
Figures~\ref{fig:finalrenders_highquality} shows results of our volumetric complex luminaires integrated in synthetic scenes; our method allows high-quality approximation of complex luminaires with a small overhead. In Figures~\ref{fig:teaser} and \ref{fig:finalrenders} we compare our method against explicitly modeling and rendering the luminaires using path tracing (PT or VPT).
Our method shows less variance and higher quality at equal time, while for equal RMSE, our method outperforms explicit PT by almost two orders of magnitude (Table \ref{table:times}), with minimal visual differences with respect to the reference. 
%

\begin{table*}[!htbp]
\centering
\begin{tabular}{ll|c|c|c|c|}
\cline{3-6}
                                                             &      & spp  & RMSE   & Time   & Speedup                       \\ \hline
\multicolumn{1}{|l|}{\textsc{Portica} Living Room}   & PT   & 4096 & 7.1211 & 84 min & 96.2 \\ \cline{2-5}
\multicolumn{1}{|l|}{}                                       & Ours & 32   & 6.8146 & 52.4 s &                               \\ \hline
\multicolumn{1}{|l|}{\textsc{Dallas} Bedroom}      & PT   &  1024    &  7.9824   & 24.6 min & 17.57             \\ \cline{2-5}
\multicolumn{1}{|l|}{}                                       & Ours &  32    &   7.9228  &  1.4 min  &                               \\ \hline
\multicolumn{1}{|l|}{\textsc{Cluster} Living Room 2} & VPT  & 1024 & 7.5267 & 19 min & 95.8 \\ \cline{2-5}
\multicolumn{1}{|l|}{}                                       & Ours & 4    & 7.4239 & 11.9 s &                               \\ \hline
\multicolumn{1}{|l|}{\textsc{Neoclassical} Bedroom}    & PT  & 1024     &  7.2304 &  14.1 min   & 38.11             \\ \cline{2-5}
\multicolumn{1}{|l|}{}                                       & Ours & 16   & 7.3782 &  22.2 s &                               \\ \hline
\multicolumn{1}{|l|}{\textsc{Star} Living Room 2}  & VPT   &  512    &  2.6439     &   7.8 min     & 8.83             \\ \cline{2-5}
\multicolumn{1}{|l|}{}                                       & Ours &   32   &  2.7113      &    53 s    &                               \\ \hline
\end{tabular}
\caption[]{\textbf{Equal-Quality Results. } We compare our method against traditional path tracing or volumetric path tracing, depending on the luminaire. At equal RMSE, our method shows significant speedups, of up to two orders of magnitude, depending on the rendering complexity of the modelled luminaire. Corresponding renders of this table can be found in Figure \ref{fig:finalrenders}}
\label{table:times}

\end{table*}

\section{Discussion \& Limitations}
\label{ch:conclusion}

\paragraph*{Comparison with Zhu et al. \cite{zhu2021neural}}
While it is difficult to compare between different systems, our model offers similar quality and speedups as neural complex luminaries, with a significantly smaller training dataset. \new{Table~\ref{table:zhu_comp} compares the reconstruction error of Zhu et al.'s method with ours in the \textsc{Dallas} dataset with the same high-resolution 100 training images. We obtain a $\times 3$ improvement in terms of PSNR. Zhu et al.'s method in general requires three to four orders of magnitude more training data for achieving comparable PSNR. } 
Rendering-wise, our model is easy to integrate in any CPU-based rendering pipeline, without the need of complex neural inference methods over a GPU\new{, and introduces a comparable penalty cost as Zhu et al's when run on a GPU. Finally, our model is two orders of magnitude heavier in terms of memory storage, although Zhu's requires a significant amount of GPU memory for running.}

\begin{table}[]
    \centering
    \begin{tabular}{l||c|c||c|}
    \cline{2-4}
                                 & PSNR           & SSIM           & Render Time (s) \\ \hline
    \multicolumn{1}{|l|}{\cite{zhu2021neural}} & 10.96          & 0.611          & 0.93                \\ \hline
    \multicolumn{1}{|l|}{Ours}   & \textbf{31.50} & \textbf{0.937} & \textbf{0.11}       \\ \hline
    \end{tabular}
    \caption[Comparison with Zhu]{\new{ \textbf{Comparison with Zhu et al.~\cite{zhu2021neural}.} Peak signal to noise ratio (PSNR) and structural similarity (SSIM) of Zhu et al.'s neural luminaires~\cite{zhu2021neural} and our method for the \textsc{Dallas} dataset (100 images, $800\times800$).  We also report rendering time on an unoptimized standalone GPU rendering implementation for both methods.}}
    \label{table:zhu_comp}
\end{table}

\new{\paragraph*{Comparison with Velazquez et al. \cite{velazquez2015complex}}
As opposed to the method by Velazquez et al., our method is able to both handle emission and appearance seamlessly, making it easier to integrate in a render engine, while providing high-quality render and illumination of the luminaires.}


\new{\paragraph*{Limitations -- Precomputation cost}
Precomputation times are rather long (2-3 days including both data generation and training), which is an issue of most complex luminaire rendering methods. While part of such cost comes from rendering the dataset, our training is still lengthy, and includes both training a NeRF and projecting it to an octree, following Plenoctrees \cite{yu2021plenoctrees}. 
A faster approach is to train directly the SH-based octree~\cite{yu2021plenoxels}, which requires a total variation regularization to avoid aliasing at discontinuities, as well as fine per-scene parameter tuning for best performance. }


\paragraph*{Limitations -- Sampling}
Currently, our sampling routine is a sub-optimal naïve surface-based sampling strategy. Investigating novel sampling techniques that take advantage of implicit learned representations of the luminaires is a promising avenue for future work.
\paragraph*{Limitations -- Illumination from other luminaires} Our method only captures the light transport from paths inside the luminaire, and ignore inter-reflections from the rest of the scene (i.e. assumes emitting-only luminaires). Accounting for paths outside the luminaire would require adding an additional term encoding the light transport response of the luminaire as a function of the incident light field.


\paragraph*{Conclusions}
In this work we have introduced a volumetric representation for complex luminaires that is both fast and capable of achieving high levels of quality. We have extended neural radiance fields to fit the particular needs of modelling complex luminaires (a high dynamic range, accurate transparency, and capable of modelling null emission, among others). Then, we have successfully exploited our volumetric representation within traditional rendering pipelines, showing significant speedups, and using significantly less input data and computational resources than previous methods. 


\paragraph*{Acknowledgements}
We want to thank the authors of \cite{zhu2021neural} for sharing their code with us. This work has been partially supported by the European Research Council (ERC) under the EU Horizon 2020 research and innovation programme (project CHAMELEON, grant No 682080), the EU MSCA-ITN programme (project PRIME, grant No 956585) and the Spanish Ministry of Science and Innovation (project PID2019-105004GB-I00). Jorge Condor acknowledges support from a grant from the I3A Institute (Beca TFM + Practicas).


\printbibliography                

\end{document}